\documentclass[preprint]{aastex}
\usepackage{epsf}
\usepackage{natbib}
\usepackage{graphicx}
\usepackage{float}
\usepackage{afterpage}
 \usepackage{relsize}
\usepackage{amsmath}
\usepackage{epsfig}
\usepackage{enumitem}
\usepackage{bm,bbm,amssymb,color, amsmath, color}
\usepackage{enumerate}
\usepackage{ulem}
\usepackage{url}
\usepackage[section]{placeins}

\begin{document}
\title{Lidov-Kozai Cycles with Gravitational Radiation: Merging Black Holes in Isolated Triple Systems}
\author{Kedron Silsbee \altaffilmark{1} \& Scott Tremaine \altaffilmark{2}}
\altaffiltext{1}{Department of Astrophysical Sciences, 
Princeton University, Ivy Lane, Princeton, NJ 08544, USA;
ksilsbee@astro.princeton.edu}
\altaffiltext{2}{Institute for Advanced Study, 
1 Einstein Drive Princeton, NJ 08540, USA;
tremaine@ias.edu}
\begin{abstract}
We show that a black-hole binary with an external companion can undergo Lidov-Kozai cycles that cause a close pericenter passage, leading to a rapid merger due to gravitational-wave emission.  This scenario occurs most often for systems in which the companion has mass comparable to the reduced mass of the binary and the companion orbit has semi-major axis within a factor of $\sim 10$ of the binary semi-major axis.  Using a simple population-synthesis model and 3-body simulations, we estimate the rate of mergers in triple black hole systems in the field to be about six per Gpc$^3$ per year in the absence of natal kicks during black hole formation.  This value is within the low end of the 90\% credible interval for the total black-hole black-hole merger rate inferred from the current LIGO results.  There are many uncertainties in these calculations, the largest of which is the unknown distribution of natal kicks.  Even modest natal kicks of $40\mbox{\, km s}^{-1}$ will reduce the merger rate by a factor of 40.  A few percent of these systems will have eccentricity greater than 0.999 when they first enter the frequency band detectable by aLIGO (above 10 Hz). 

 \end{abstract}
\section{Introduction}
The announcement of a black-hole merger event detected in gravitational waves by the advanced LIGO (aLIGO) experiment \citep{Abbott16a} has highlighted the question of how to form black hole-black hole binaries that can merge in less than the age of the universe via gravitational-wave emission.
\par 
The merger time for two black holes on a circular orbit due to gravitational-wave emission is 
\begin{equation}
\label{tauMerge}
\tau_{\rm merge} = \frac{5}{256} \frac{c^5}{G^3} \frac{a^4}{M^2 \mu} = 10^{10} \, {\rm years}\left(\frac{20 M_\odot}{M}\right)^2 \frac{5M_{\odot}}{\mu} \left(\frac{a}{0.0888\, {\rm AU}}\right)^4
\end{equation}
\citep{Peters64}.  Here $a$ is the semi-major axis of the black-hole binary, $M = m_1 + m_2$ is the sum of the individual masses, and $\mu = m_1 m_2/(m_1 + m_2)$ is the reduced mass of the system.  Equation \eqref{tauMerge} implies that a binary with black-hole masses similar to those in the detected merger GW150914 ($M = 65M_\odot$, $\mu = 16 M_\odot$) will merge in less than the age of the universe if the separation is less than about 0.23 AU.  This separation is smaller than the maximum radius of an O-star in the later stages of stellar evolution (which we estimate to be between 0.3 and 5 AU depending on mass, from the luminosity-temperature diagrams shown in Figure 5 of \citet{Meynet03}; see Table \ref{table:StellarRadii}), raising the question of how such a black-hole binary would form.  This tension between the minimum orbital radius of the binary and the maximum radii of its components is enhanced by the outward expansion of the orbit due to mass loss that occurs after the time of maximum radius.  For these reasons, the standard picture of the merger of an isolated black-hole binary involves a common-envelope phase, which dramatically shrinks the orbit of the stellar cores, before they collapse to black holes \citep{Belczynski02, Dominik12}.  Another proposed model is ``chemically homogeneous evolution" in which the stars in the binary remain almost fully mixed through their lifetimes, and consequently do not inflate dramatically at the end of their lives \citep{Mandel16b}.
\par
Black-hole binaries capable of merging in less than the age of the universe could also form dynamically in globular clusters \citep{Rodriguez16a}.  Total estimated rates for mergers occurring in these clusters are considerable: \citet{Rodriguez16a} [erratum in \citet{Rodriguez16b}] estimate that aLIGO should detect 30 mergers per year from black-hole binaries assembled in globular clusters.  
\par
 \citet{Antonini16} study black holes that merge due to evolution in triple systems that dynamically form in globular clusters.  They estimate that this mechanism could produce on the order of one event per year detectable with aLIGO.  While these rates are substantially smaller than the total rate for all black-hole binaries formed in globular clusters \citep{Rodriguez16a}, these mergers are particularly interesting because, in contrast to mergers due to other channels, approximately 20\% of these merging binaries would have eccentricities greater than 0.1 when they first enter the aLIGO detection band (greater than 10 Hz), and approximately 10\% would have eccentricity greater than 0.9999.
\par
In this paper, we describe an alternative formation channel based on orbital evolution in an isolated triple system containing a black-hole binary and an external companion.  Perturbations from the external companion can cause the orbit of the inner binary to become nearly radial, via the Lidov-Kozai mechanism (see \citet{Naoz16} for a review).  Energy is then lost to gravitational radiation, which merges the inner orbit faster than the outer companion can change the pericenter of the inner orbit.  We show via N-body simulations that this mechanism causes the inner binary to merge within 10 Gyr in a few percent of moderately hierarchical triple systems containing three black holes.
\section{Analytic Estimates of System Parameters}
\label{analyticEstimate}
\subsection{Critical pericenter distance necessary to decouple the inner binary from the companion star}
In this section we estimate the critical pericenter distance for the inner binary such that the orbital decay induced by the emission of gravitational waves during one pericenter passage is sufficient to shrink the orbit enough that the pericenter distance is immune to further changes caused by the perturber.  Provided that this shrinkage corresponds to a change in $a_{\rm in}$ (the semi-major axis of the inner binary) that is at least of order unity, the orbit will continue to shrink faster than it can be torqued out of its nearly radial configuration.  
\par
Let the current separation between the two inner black holes be $R_{\rm in}$.  For simplicity, in this section we assume that the outer orbit is circular.  Let the separation between the barycenter of the inner binary and the outer star be $a_{\rm out}$.  Let the masses of the inner black holes be $m_1$ and $m_2$, and the mass of the outer companion be $m_3$, and define $M_{\rm in} = m_1 + m_2$, $\mu = m_1m_2/(m_1 + m_2)$, and $M_{\rm tot} = m_1 + m_2 + m_3$.  Let the angle between the line passing through the inner stars and the line through the barycenter of the inner stars and the outer star be $\theta$.  Then, in the quadrupole approximation, valid when $a_{\rm out} \gg R_{\rm in}$, the instantaneous torque $\Gamma$ on the inner orbit due to the outer star is
\begin{equation}
\label{gamma}
\Gamma = \frac{3Gm_3\mu R_{\rm in}^2 \sin{2\theta}}{2a_{\rm out}^3},
\end{equation}
where $G$ is Newton's constant.  Assuming the period of the outer orbit to be much longer than the period of the inner orbit, and the inner orbit to be almost radial (so $\theta$ is constant over one orbit of the inner binary), we can average this torque over the inner orbit (while holding the outer star at fixed position).  Under these assumptions, 
\begin{equation}
\langle R_{\rm in}^2 \rangle = \frac{5}{2} a_{\rm in}^2.
\end{equation} 
Then the average torque is given by 
\begin{equation}
\label{gammaav}
\langle \Gamma \rangle = \frac{15Gm_3\mu a_{\rm in}^2 \sin{2\theta}}{4a_{\rm out}^3}. 
\end{equation}
Since we are assuming that the orbit is almost radial, the angular momentum of the inner orbit, $L_{\rm in}$, is
\begin{equation}
\label{qofL}
L_{\rm in} = \mu \sqrt{2GM_{\rm in} q_{\rm in}},
\end{equation}
where $q_{\rm in}$ is the pericenter distance of the inner orbit.
\par
Let $P_{\rm in} = 2 \pi a_{\rm in}^{3/2}/(GM_{\rm in})^{1/2}$ be the period of the inner binary, and let $\Delta L_{\rm in} = \langle \Gamma \rangle P_{\rm in}$ be the magnitude of the change in angular momentum accrued over one orbit of the inner binary.  Normalized by $L_{\rm in}$, this is 
\begin{equation}
\label{deltal}
 \frac{\Delta L_{\rm in}}{L_{\rm in}} = \frac{15 \pi \sqrt{2}}{4} \left(\frac{a_{\rm in}}{a_{\rm out}}\right)^3 \left(\frac{a_{\rm in}}{q_{\rm in}}\right)^\frac{1}{2} \frac{m_3}{M_{\rm in}} \sin{2\theta}.
 \end{equation}
\par
The orbit-averaged rate of change of the period $P_{\rm in}$ of a binary star system with masses $m_1$ and $m_2$ due to gravitational radiation is \citep[e.g.,][]{Shapiro83}
\begin{equation}
\label{eq:Pin}
\frac{1}{P_{\rm in}} \frac{dP_{\rm in}}{dt} = -\frac{96}{5} \frac{G^3 M_{\rm in}^2 \mu}{a_{\rm in}^4 c^5} f(e_{\rm in}),
\end{equation}
where $e_{\rm in}$ is the eccentricity, and 
\begin{equation}
\label{fe}
f(e_{\rm in}) = \frac{1 + \frac{73}{24} e_{\rm in}^2 + \frac{37}{96} e_{\rm in}^4}{(1-e_{\rm in}^2)^{7/2}}.
\end{equation}
Equation \eqref{eq:Pin} is not so useful as it stands, because for nearly parabolic orbits, $a_{\rm in}$, $P_{\rm in}$, and $1-e_{\rm in}$ can change substantially during one pericenter passage.  However, $q_{\rm in}$ and the energy $E = -GM\mu/(2a_{\rm in})$ do not change much.  For this reason, we re-write the equation in terms of the change in energy in one pericenter passage as a function of pericenter distance, assuming $e \sim 1$:
\begin{equation}
\label{DeltaE}
\Delta E = -\frac{85 \pi \sqrt{2}}{24} \frac{G^{7/2}M_{\rm in}^{5/2}\mu^2}{q_{\rm in}^{7/2} c^5},
\end{equation}
where we have used the approximation
\begin{equation}
\label{fapprox}
f(e_{\rm in}) \approx \frac{425 \sqrt{2}}{1536} \left(1-e_{\rm in}\right)^{-7/2},
\end{equation}
valid as $e_{\rm in} \rightarrow 1$.
\par
 Let $a_1$ and $a_2$ be the values of $a_{\rm in}$ before and after a pericenter passage $P$.  Let $L_1$ be the angular momentum of the inner orbit at pericenter passage $P-1$, and let $\Delta L_1$ be the change in the angular momentum of the inner orbit between passages $P-1$ and $P$.  Let $L_2$ and $\Delta L_2$ be the angular momentum of the inner orbit at passage $P$, and the change in the angular momentum of the inner orbit between passages $P$ and $P+1$.  Let us assume further that $\Delta L_1/L_1 \gg 1$.
\par
In order for further inspiral to occur after $P$, we require that $a_2$ be small enough that the orbit cannot be misaligned from its radial configuration prior to the next pericenter passage, i.e., $\Delta L_2/L_2 \ll 1$.  Since we have assumed $\Delta L_1/L_1 \gg 1$, in order to have $\Delta L_2/L_2 \ll 1$, we require that $a_2 \ll a_1$.  Therefore, the initial binding energy prior to $P$ can be ignored, and we can solve for $a_2$ using Equation \eqref{DeltaE}:
\begin{equation}
\label{ain}
a_2=\frac{6\sqrt{2}}{85\pi}\frac{q_1^{7/2}c^5}{G^{5/2}M_{\rm in}^{3/2}\mu}.
\end{equation}
 Plugging Equation \eqref{ain} into Equation \eqref{deltal}, and requiring that $|\Delta L_2| < L_2$, we find that $q_{\rm in} < q_{\rm crit}$, where
\begin{align}
\label{qcrit}
q_{\rm crit}& = 2.20\frac{G^{35/47} M_{\rm in}^{25/47} \mu^{{14/47}} a_{\rm out}^{12/47}}{c^{70/47} m_3^{4/47}|\sin{2\theta}|^{4/47}} \nonumber\\
&= 1.28\cdot 10^4\; {\rm km} \left(\frac{M_{\rm in}}{40 M_\odot}\right)^{25/47} \left(\frac{\mu}{10 M_\odot}\right)^{14/47} \left(\frac{a_{\rm out}}{100 AU}\right)^{12/47}\left(\frac{20M_\sun}{m_3}\right)^{4/47},
\end{align}
We have let $\sin{2 \theta} = 1$ in the numerical estimate.  
\par
In order to check that the assumption $\Delta L_1/L_1 \gg 1$ is valid, we substitute $q_{\rm crit}$ from Equation \eqref{qcrit} into Equation \eqref{deltal}.  The requirement that $\Delta L_1/L_1 > 1$ is only met if
\begin{align}
& \frac{a_{\rm out}}{a_{\rm in}} < \frac{2.167 a_{\rm in}^{35/294} c^{70/294}m_3^{98/294}}{\mu^{14/294}M_{\rm in}^{119/294}G^{35/294}}  \nonumber\\
= &15.9 \left(\frac{a_{\rm in}}{30 \, {\rm AU}}\right)^{35/294} \left(\frac{m_3}{20M_\odot}\right)^{98/294} \left(\frac{20 M_{\odot}}{\mu}\right)^{14/294} \left(\frac{40 M_\odot}{M_{\rm in}}\right)^{119/294}.
\end{align}

This separation ratio is larger than, but comparable to the upper cut-off to the separation ratio distribution derived in the following section, showing that the approximation $\Delta L_{\rm in}/L_{\rm in} \gg 1$ used to derive Equation \eqref{qcrit} is plausible, but likely to lead to some error.  We show in simulations below that we underestimate $q_{\rm crit}$ by about a factor of two, with a few cases achieving an inspiral while maintaining even higher values of $q_{\rm in}$ (see Figure \ref{RratioHist}).

\subsection{Critical separation ratio such that the Lidov-Kozai resonance is not destroyed by apsidal precession}
\label{kozai}
Lidov-Kozai oscillations are quenched by the relativistic precession of the inner binary unless \citep{Blaes02}
\begin{equation}
\label{hierRat1}
\left(\frac{a_{\rm out}}{a_{\rm in}}\right)^3 <  \frac{3c^2m_3a_{\rm in} (1-e_{\rm in}^2)^{3/2}}{4GM_{\rm in}^2(1-e_{\rm out}^2)^{3/2}} \rightarrow \frac{a_{\rm out}}{a_{\rm in}} < 305.6 \left(\frac{a_{\rm in}}{30\, {\rm AU}}\right)^{1/3} \left(\frac{M_{\rm in}}{40M_\odot}\right)^{-1/3} \left(\frac{2m_3}{M_{\rm in}}\right)^{1/3} \sqrt{\frac{1-e_{\rm in}^2}{1-e_{\rm out}^2}}.
\end{equation}
Here $e_{\rm out}$ is the eccentricity of the outer binary.  Equation \eqref{hierRat1} considers only quadrupole-order terms in the disturbing function.  As stated before, we are assuming $e_{\rm out} \approx 0$.  We cannot, however, assume that $e_{\rm in}\approx 0$, since we are interested in the behavior precisely when $e_{\rm in} \approx 1$. 
\par
As is stated in \citet{Antonini12} and \citet{Katz12}, the inner binary is actually able to explore more of phase space than it does in the double-averaged approximation used to derive Equation \eqref{hierRat1} because of variations in the inner orbit on the timescale of the outer orbital period.  It seems plausible that deviations in $L_{\rm in}$ from the secular approximation will be on the order of the amount of $L_{\rm in}$ that can be accumulated over a quarter of the orbit of the outer binary (the time over which $\Gamma$ keeps the same sign) \citep{Ivanov05}. Using Equation \eqref{gamma}, we find that this variation is given by
\begin{equation}
\label{DeltaL}
\Delta L_{\rm dev} = \frac{15 \pi}{8} \frac{G^{\frac{1}{2}} m_3 \mu a_{\rm in}^2}{M_{\rm tot}^{\frac{1}{2}} a_{\rm out}^{\frac{3}{2}}}.
\end{equation}
To determine how high $e_{\rm in}$ must be so that this ``wobble" could take $L_{\rm in}$ to 0, we set $L_{\rm in}$ in Equation \eqref{qofL} equal to $\Delta L_{\rm dev}$ in Equation \eqref{DeltaL} and solve for $q_{\rm close}$, the value of $q_{\rm in}$ such that the $\Delta L_{\rm dev}$ accrued over a quarter orbit of the outer binary is equal to $L_{\rm in}$:
\begin{equation}
\label{qrmin}
q_{\rm close} = \frac{225 \pi^2}{128} \frac{m_3^2 a_{\rm in}^4}{M_{\rm in} M_{\rm tot} a_{\rm out}^3}.
\end{equation}
Using the relation $1-e_{\rm in}^2 \approx 2q_{\rm in}/a_{\rm in}$, we can re-write Equation \eqref{hierRat1} as
\begin{equation}
\label{hierRatInt}
\mathlarger{\frac{a_{\rm out}}{a_{\rm in}}}  < 432.2 \left(\frac{a_{\rm in}}{{30 \, \rm AU}}\right)^{\frac{1}{3}} \left(\frac{M_{\rm in}}{40 M_{\odot}}\right)^{-\frac{1}{3}} \left(\frac{2m_3}{M_{\rm in}}\right)^\frac{1}{3} \left(\frac{q_{\rm in}}{a_{\rm in}}\right)^\frac{1}{2}. 
 \end{equation}
 We are using Equation \eqref{qrmin} to derive Equation \eqref{hierRatInt} instead of Equation \eqref{qcrit} because we expect $q_{\rm close} > q_{\rm crit}$ for the following reasons.  We concluded in the previous section that for our fiducial parameters, $\Delta L_{\rm in}/L_{\rm in} > 1$ at $q_{\rm in} = q_{\rm crit}$, where $\Delta L_{\rm in}$ is the change in angular momentum of the inner orbit accrued over one inner orbit.  Since $1/4$ of an outer orbit takes longer than one inner orbit, $\Delta L_{\rm dev} > \Delta L_{\rm in}$ for a given value of $q_{\rm in}$.  Therefore, $q_{\rm close} > q_{\rm crit}$.
\par
 Then replacing $q_{\rm in}$ by $q_{\rm close}$ from Equation \eqref{qrmin} and solving, we find
\begin{equation}
\label{hierRat2}
\frac{a_{\rm out}}{a_{\rm in}} < 14.0 \left(\frac{a_{\rm in}}{30\rm AU}\right)^\frac{2}{15} \left(\frac{M}{40 M_\odot}\right)^{-\frac{2}{15}} \left(\frac{6m_3^2}{M_{\rm in} M_{\rm tot}}\right)^\frac{1}{5}\left(\frac{2m_3}{M_{\rm in}}\right)^{\frac{2}{15}}.
\end{equation}
The key assumption in the derivation of Equation \eqref{hierRat2} is that averaging over both orbits (used to derive Equation \eqref{hierRat1}) is a good approximation until the pericenter distance is less than $q_{\rm close}$.  At this point, changes to the inner orbit on the timescale of the outer orbit can bring the pericenter distance to zero.  We have calculated $q_{\rm close}$ assuming that the only quantity that changes appreciably in the inner orbit on the timescale of the outer orbital period is the pericenter distance.  To ensure that this is valid, we should check that the orbital precession of the inner binary due to general relativity is less rapid than the mean motion of the outer binary.  Using the formula for orbital precession due to general relativity given in \citet{MTW}, we find that this condition is satisfied as long as 
\begin{equation}
\frac{P_{\rm in}}{P_{\rm out}} > \frac{3R_{\rm EH}}{q_{\rm in}},
\end{equation}
where $R_{\rm EH} = 2GM_{\rm in}/c^2$ is the radius of the event horizon of a black hole with the mass of the inner binary.  We see from Equations \eqref{qcrit} and \eqref{hierRat2} that this condition is satisfied for typical systems at $q_{\rm in} = q_{\rm crit}$ but not with a large margin of error.
\subsection{Minimum separation necessary for stability}
Even ignoring the issue of mergers due to Lidov-Kozai oscillations, there is also a minimum separation of the inner and outer orbits necessary for stability.  Define 
\begin{equation}
\label{eq:Y}
Y = \frac{a_{\rm out}(1-e_{\rm out})}{a_{\rm in}(1+e_{\rm in})};
\end{equation}
$Y$ is the ratio of the periastron distance of the outer orbit to the apoastron distance of the inner orbit.  \citet{Kiseleva96} give a critical value of $Y$, which roughly divides stable from unstable systems:
\begin{equation}
\label{eq:Ycrit}
Y_{\rm crit} = \frac{3.7}{Q_{\rm out}} - \frac{2.2}{1+Q_{\rm out}} + \frac{1.4}{Q_{\rm in}}\, \,\frac{Q_{\rm out} - 1}{Q_{\rm out} + 1},
\end{equation}
where $Q_{\rm in} = \left[{\rm max}(m_1, m_2)/{\rm min}(m_1, m_2)\right]^{1/3}$, and $Q_{\rm out} = \left[(m_1 + m_2)/m_3\right]^{1/3}$.  
For a system in which all three components have equal mass, and both orbits have zero eccentricity, $Y_{\rm crit} = 2.1$.  Equations \eqref{hierRat2} through \eqref{eq:Ycrit} imply that there is a rather narrow window of acceptable semi-major axis ratios that are large enough to permit stability, but small enough so that the Lidov-Kozai cycles still raise the eccentricity to very high values despite the relativistic precession.  
\par
Our simulations confirm that these estimates are reasonable.  We find that the majority of merging systems have $Y/Y_{\rm crit}$ within a factor of three of the stability boundary (see Figure \ref{outcomes}).  In the circular case with equal masses, $Y/Y_{\rm crit} = 3$ corresponds to $a_{\rm out}/a_{\rm in} =  6.3$.  

\section{Simulation Set-up \& Input Parameters}
\label{setup}
In order to get a more accurate estimate of the fraction of triple systems that lead to merger, we ran 3-body simulations using the Rebound package \citep{Rein12} with the IAS15 integrator \citep{Rein15}.  We modified the code to include two general relativistic effects --- apsidal precession, and gravitational-wave emission by the inner binary.  
\par
Precession due to general relativity was implemented by adding a term 
\begin{equation}
\label{eq:UGR}
U_{\rm GR} = -\frac{3G^2m_im_j(m_i + m_j)}{c^2R_{ij}^2}
\end{equation}
to the potential energy between bodies $i$ and $j$, where $R_{ij}$ is the distance between those bodies \citep{Wegg12}.  This potential reproduces the correct apsidal precession rate at all eccentricities as long as $R_{ij} \gg 2G(m_i+m_j)/c^2$.
\par
To reproduce the energy loss from gravitational radiation, we used Equation 16.2.16 from \citet{Shapiro83}, which states that the gravitation wave luminosity is
\begin{equation}
L_{\rm GW} = -\frac{dE}{dt} = \frac{1}{5} \frac{G}{c^5}\langle \dddot  I_{jk} \dddot I_{jk}\rangle,
\end{equation}
where $I_{jk}$ is the quadrupole moment of the inner binary, given by 
\begin{equation}
\label{Ijk}
I_{jk} = \sum_A m_A \left[x_j^Ax_k^A - \frac{1}{3} \delta_{jk}(x^A)^2\right].
\end{equation}
and the angle brackets denote a time average.  In Equation \eqref{Ijk}, $A = 1, 2$ denotes the two components of the binary and the origin of the coordinates is at the center of mass.  We dropped the angle brackets and applied a force opposing the direction of motion with the right magnitude to reproduce the correct energy loss.  This procedure is not exactly correct, but gives the correct energy loss averaged over one orbit, and concentrates the energy loss at pericenter as required\footnote{The correct formula for the instantaneous reaction force is given by $F^{({\rm react})} = -m \nabla \Phi^{({\rm react})}$, where $\Phi^{({\rm react})} = GI^{(5)}_{jk} x_j x_k/(5c^5)$.  $I^{(5)}_{jk}$ denotes the 5$^{\rm th}$ time derivative of $I_{jk}$. }.  Essentially, both our approximate scheme and the correct physics give a kick at pericenter that changes the energy of the orbit while having very little effect on the angular momentum. Our approximation does not correctly account for the angular momentum lost to gravitational radiation, but this is only important when the binary has lost so much energy that it will inevitably merge in a short time.  We expect gravitational radiation by the outer binary to be negligible and do not consider it in our simulations.  More discussion of our implementation of general relativistic effects is provided in Appendix \ref{verification}.
\subsection{Picking the stellar masses}
\label{masses}
In this section we describe our algorithm for assigning the masses to our stellar systems.  If the third body is a white dwarf or neutron star remnant, it will generally not be massive enough to make the inner orbit completely radial, since the outer orbit will have less angular momentum than the inner orbit unless the eccentricity of the inner binary is nearly one, or the system is too hierarchical for the Lidov-Kozai mechanism to operate given the relativistic apsidal precession of the inner orbit.  We therefore only simulate systems in which we believe all three components will become black holes.  We take a simple model in which a star becomes a black hole if and only if its zero age main sequence (ZAMS) mass exceeds $25 M_\odot$.   
\par
Since 77/136 of our modeled systems merge in less than the lifetime of an 8 $M_\odot$ star \citep[40 Myr;][]{Schaller92}, it is possible that a main-sequence companion of several solar masses could also play the role of the external companion.  We do not investigate this possibility further, but note that inclusion of main-sequence companion stars could increase the merger rate.
\par
We used the following algorithm to assign masses to the stars in our simulation.  We drew the first star in the inner binary from a Kroupa IMF with an upper cut-off of 150 $M_\odot$ \citep{Kroupa01}.  We took the local star formation rate density to be $0.025 M_\odot \, {\rm yr}^{-1} \, {\rm Mpc}^{-3}$ \citep{Bothwell11}.  This means that the number of stars of mass $m>0.5$ (in solar masses) formed per unit mass per year per Gpc$^3$ is given by 
\begin{equation}
\label{NofM}
N_m(m)dm = 5.40\cdot 10^{6} m^{-2.3}.
\end{equation}
\par
  We next assign a mass to the second star in the inner binary.  \citet{Sana13} find that the distribution of the mass ratio $\kappa$ in massive binaries is approximately log-uniform on the range $0.1 < \kappa < 1$, i.e.,
\begin{equation}
\label{Nqm}
N_\kappa(\kappa)d\kappa = \frac{\kappa^{-1}}{\ln{10}}d\kappa; \quad 0.1 < \kappa < 1,
\end{equation}
and zero otherwise.  We pick a value of $\kappa_{\rm in}$ for the inner binary from Equation \eqref{Nqm}.  The mass $m_2$ of the second star is thus limited to either $\kappa_{\rm in} m_1$ or $m_1/\kappa_{\rm in}$, depending on which star is larger.  To assign probabilities to each case, we made the assumption that Equation \eqref{Nqm} is independent of the mass of the primary star and the mass distribution of stars that are formed in binary systems with mass ratio $\kappa_{\rm in}$ is the same as the IMF.\footnote{This method of mass assignment is different from the commonly used method in which the most massive star is picked from the IMF and the less massive star is picked from the distribution of mass ratios.  This standard method leads to the mass distribution of the binary population being skewed towards lower masses than the IMF.  We discuss the effect of using the standard method of picking masses in section \ref{otherModels}.}  Given the above assumptions, we show in Appendix \ref{binSelecAlg} that if a randomly selected star in a binary has mass $m_1$, then the probability that the second star is more massive than the first is given by $\alpha(m_1, \kappa_{\rm in})$, where
\begin{equation}
\label{alphaExact}
\alpha(m_1, \kappa_{\rm in}) = \Sigma_{i = 0}^{L-1} (-1)^i \kappa_{\rm in}^{1.3\cdot(1+i)},
\end{equation}
\begin{equation}
\label{leveleq}
L = \left \lfloor \frac{\ln{m_1} - \ln{m_{\rm max}}}{\ln{\kappa_{\rm in}}}\right \rfloor,
\end{equation}
and $\lfloor x \rfloor$ denotes the largest integer smaller than $x$.  Let $\kappa_{\rm out}$ be the mass ratio of the outer binary, where the mean mass of the two inner components is used as the other mass in defining $\kappa_{\rm out}$.  We assume that $\kappa_{\rm out}$ is also drawn from Equation \eqref{Nqm}, and that the distribution in Equation \eqref{Nqm} is again independent of the masses.  We then pick a value of $m_3$ given $m_1$ and $m_2$ following the same procedure we used to pick $m_2$ given $m_1$.  Note that we will sometimes pick $m_2$ or $m_3 < 25 M_\odot$.  In this case we discard the system and start over from the beginning of the process. 
\par
Figure 13 of \citet{Sana14} suggests that 81\% of O-stars in their magnitude-limited sample have at least one companion either spectroscopically detected, or resolved within a projected separation of about 6,000 AU.  Both detection methods were able to identify companions down to about 10\% of the primary mass.  About 25\% have at least two such companions.  We assume that the same statistics hold for the population of stars with $M > 25 M_\odot$ as for the population of O-stars, even though the latter category is slightly more inclusive, as O-star masses can be as small as 16 $M_\odot$ \citep{Martins05}.
\par
As discussed above, we only simulate systems in which all three stars are black-hole progenitors (have ZAMS masses of at least 25 $M_\odot$).  We therefore wish to determine the ratio $\epsilon_{\rm trip}$ of black-hole progenitor triples to total black-hole progenitors.  In this section, we ignore systematic effects associated with the \citet{Sana14} survey (the sample was magnitude-limited, and therefore biased towards high masses and systems with multiple unresolved O-stars), and assume the following: 
\begin{itemize}
\item {All black-hole progenitors are in single, double, or triple systems (we ignore higher-multiplicity systems).}
\item The mass ratios in multiple systems are as described above.
\item{19\% of stellar systems with at least one black-hole progenitor are single systems, 56\% are binary systems, and 25\% are triple systems: these numbers come from \citet{Sana14} as described above.}
\end{itemize}
\par
Then, there are six different types of systems: ``B", ``BN", ``BB", ``BNN", ``BBN", and ``BBB", where ``B" denotes a black-hole progenitor, and ``N" denotes a less massive star.  Using the model for the masses of different components of multiple systems described in this section, we find that the ratio of ``BB" systems to ``BN" systems is 35 to 65, and the ratio of ``BNN" to ``BBN" to ``BBB" systems is 51:35:14.  Therefore, in a sample of 100 stellar systems with at least one black-hole progenitor, there are an expected $25\cdot 0.14$ = 3.5 ``BBB" systems, and a total of $19 + 56 (0.65 + 2\cdot 0.35) + 25(0.51 + 2 \cdot 0.35 + 3 \cdot 0.14)$ = 135.35 black-hole progenitors.  Therefore, given these approximations, $\epsilon_{\rm trip} = 3.5/135.5 = 0.026$.  We remind the reader that as discussed above, it is possible that some ``BBN" systems could also yield a merger.  We are providing a conservative estimate by not considering this possibility.
\par

\subsection{Orbital elements}
\label{orbitalsma}
We characterize the strength of the interaction between the inner and the outer binary by  what we call the stability ratio:
\begin{equation}
\label{Rs}
R_s = \frac{Y}{Y_{\rm crit}}
\end{equation}
where $Y$ is given by Equation \eqref{eq:Y} and $Y_{\rm crit}$ by Equation \eqref{eq:Ycrit}.  A system is stable roughly if $R_s > 1$.
We pick $a_{\rm in}$ and $a_{\rm out}$ independently from a log-uniform distribution from 0.1 to 6,000 AU.  The log-uniform distribution for binary orbits (known as {\"O}pik's law) is roughly consistent with the results of \citet{Kobulnicky14}, derived from observations of 48 massive star systems in the Cygnus OB2 association, however the assumption that the inner and outer orbit in a triple system can be drawn independently from such a distribution is pure speculation.  We picked eccentricity uniform from 0.0 to 0.8 for both the inner and outer binary orbits.  This distribution is roughly consistent with the curve shown in Figure 3 of the review by \citet{Duchene13}, showing results from the literature for the eccentricities of massive stellar binaries.  We also discarded systems with $R_s < 1.0$ as being unlikely to be stable, and systems with $R_s > 12.0$ as unlikely to merge (see Section \ref{kozai}).  We picked the remaining eight angular orbital elements from an isotropic distribution. 
\par
 We define $\epsilon_{\rm suit}$ to be the fraction of systems that satisfy a set of conditions that make them suitable candidates for eventual merger: (i) initial stability ratio satisfying $1.0 < R_s < 12.0$; (ii) stability ratio $R_s>1.0$ at all times during the stellar evolution, (iii) pericenter distance of the inner and outer binary large enough that no common envelope phase occurs. Assigning masses and semi-major axes as discussed above, we find that $\epsilon_{\rm suit} = 15$\%. 
\subsection{Mass loss and black-hole natal velocities}
\label{massLoss}
The final stages of stellar evolution are quite complicated, and depend on the metallicity and rotation of the stars.  In this paper, we adopt a simple model in which each star symmetrically loses $2/3$ of its mass over a timescale long compared to the orbital timescales.  We neglect mass loss associated with the collapse to a black hole.  Mass loss causes the outer orbit to expand if the star is a member of the outer binary, and both orbits to expand if the star is a member of the inner binary.  In a binary with two stars of mass $m_a$ and $m_b$, if star $a$ adiabatically loses mass equal to $\Delta m_a$, the orbit expands by a factor of $(m_a + m_b)/(m_a + m_b - \Delta m_a)$, but the eccentricity is left unchanged.
\par
We do not expect the distribution of the angular orbital elements to change during mass loss, so long as the mass loss is slow and isotropic.  At the end of its life each star collapses to a black hole.  It is unknown how much of a velocity kick the collapse process imparts to the newly formed black hole.  In this paper, we assume that each black hole receives a kick drawn from a multivariate Gaussian distribution with three-dimensional RMS velocity $v_{\rm rms}$, i.e.,
\begin{equation}
\label{sigmakdist}
\rho(\vec v) = \frac{3 \sqrt{3}}{v_{\rm rms}^3 (2 \pi)^{3/2}}\exp{\left[{\mathlarger{-3\frac{v_x^2 + v_y^2 + v_z^2}{2v_{\rm rms}^2}}}\right]}
\end{equation}
Kicks are discussed in more detail in Section \ref{kicks}.
\par
We assume that each star undergoes expansion, mass loss and conversion to a black hole sequentially (mass loss does not begin until the maximum radius is reached), and in order from most massive to least massive --- i.e., the most massive star receives its natal kick before the second most massive star begins mass loss.  The orbital elements of the inner and outer orbit are updated as appropriate to reflect the adiabatic mass loss from one component, and then an impulsive kick.  

\begin{table}[ht]
\caption{Maximum radius as a function of ZAMS mass}
\centering
\vspace{1mm}
\begin{tabular}{c c}
\hline \hline
ZAMS Mass $(M_\odot)$ & $R_{\rm max}$ $({\rm AU})$ \\ [0.5ex] 
\hline
25 & 4.85  \\
40 & 3.36  \\
60 & 0.313  \\
85 & 0.276 \\
120 & 0.343 \\[1ex]
\hline
\label{table:StellarRadii}
\end{tabular}
\end{table}

\par
It is believed that a substantial fraction of binaries undergo a common envelope phase that dramatically shrinks the orbits.  While the details of common envelope evolution are uncertain, it seems likely that any common envelope evolution would disrupt the mechanism proposed in this paper by shrinking the inner orbit so that the system is too hierarchical for the Lidov-Kozai mechanism to operate.  For this reason, we discard a system if at any time the physical radius of one of the stars in the inner binary is larger than the pericenter distance of the inner binary, or the radius of the outer companion is larger than the pericenter of the outer orbit.  We assume that each star reaches its largest physical radius prior to beginning mass loss.  We estimate the maximum physical radius as a function of ZAMS mass by interpolating between the luminosity-temperature diagrams in the rotating models in Figure 5 of \citet{Meynet03}.  The maximum radii we estimated are presented in Table \ref{table:StellarRadii}.  We do not include the effects of mass transfer in our modeling.

\subsection{Simulation termination conditions}
\label{termination}
It was shown in \citet{Katz12} that if a triple system is going to merge, the time it will take to do so is typically no more than a few times the time it would take the inner binary to collide if it sampled a random point in the energetically accessible phase space at each pericenter passage.  This time is
\begin{equation}
\label{trand}
t_{\rm rand} = \frac{P_{\rm in}a_{\rm in}}{2q_{\rm crit}}.
\end{equation}
In the case of colliding white dwarfs discussed in \citet{Katz12}, $q_{\rm crit}$ is given by the physical collision cross-section of the white dwarfs.  In our case, it is given by Equation \eqref{qcrit}.  We run our simulations for a maximum of
\begin{equation}
\label{tsim}
\tau_{\rm sim} = {\rm min}(10 t_{\rm rand}, \, {\rm 10 \,Gyr}).
\end{equation}
We believe that this simulation time captures most, but not all of the mergers that will occur in 10 Gyr.  We discuss this issue further in Section \ref{sect:dtd}.
\par
We also terminate our simulations if the inner binary merges or if the system splits.  We consider a system to have merged if at any time step $E_{\rm in}/\Phi_{\rm in}(0.1\, {\rm AU}) > 1$, where $E_{\rm in}$ is the total energy of the inner orbit, and $\Phi_{\rm in}(0.1\, {\rm AU})$ is the potential energy of the inner binary at a separation of $0.1$ AU.  This condition is similar to removing systems with apocenter distance less than $0.1$ AU.  For all of our systems, an apocenter distance of $0.1$ AU represents shrinkage of the inner orbit by a factor of at least eight, which we assume to be sufficient to decouple the inner from the outer orbit and guarantee rapid merger.  We consider a system to have ``split" if one or more of the stars moved outside the simulation region, a cube with side length $10^5$ AU centered on the barycenter.
\section{Simulation Results and Rate Estimates}
\label{simulationResults}
\subsection{Merger rate estimate in the absence of natal kicks}
In our simulations with zero kick velocity, $\epsilon_{\rm merge} = 2.7\%$ of systems that we simulated merged within $\tau_{\rm sim}$ defined by Equation \eqref{tsim}.
\par
Therefore, in a universe with constant star formation rate per comoving volume unit, the number of black-hole merger events per Gpc$^3$ per year is 
\begin{equation}
\label{rateEquation}
N_{\rm merge} = N_{25,150} \epsilon_{\rm trip} \epsilon_{\rm suit} \epsilon_{\rm merge}  = 6.1,
\end{equation}
where $N_{25,150} = 5.71\cdot 10^4$, determined by integrating Equation \eqref{NofM}, is the current rate of formation of stars in the mass range $25M_\odot$ -- $150M_\odot$ per Gpc$^3$ per year.  $\epsilon_{\rm trip}$ and $\epsilon_{\rm suit}$ are defined in Sections \ref{masses} and \ref{orbitalsma} respectively.  
\par
 Of course, the uncertainties in these numbers are dominated by the uncertain assumptions that go into our model. These are discussed further in Section 6.
\FloatBarrier
\subsection{Horizon distance estimate and detection rates}
To estimate the horizon distance $D_H$ (the distance to which a merger with a given set of properties could be detected by the current aLIGO detectors), we used Equation (6) from \citet{LIGO12}.  This states
\begin{equation}
\label{Horizon}
D_H = \frac{1}{8} \left(\frac{5\pi}{6c^3}\right)^{1/2} \left(GM_{\rm chirp}\right)^{5/6} \pi^{-7/6} \sqrt{\int_{f_{\rm low}}^{f_{\rm high}} \frac{f^{-7/3}}{S_n(f)} df},
\end{equation}
where $f$ is the gravitational-wave frequency, $f_{\rm low}$ is set by the detector noise cut-off, $f_{\rm high} = c^3/(6\sqrt{6}\pi G M_{\rm in})$ is the frequency of the innermost stable circular orbit, $S_n(f)$ is the power spectral density of the noise, and $M_{\rm chirp}$ is the chirp mass, given by
\begin{equation}
\label{Mchirp}
M_{\rm chirp} = \frac{\left(m_1 m_2\right)^{3/5}}{\left(m_1 + m_2\right)^{1/5}}.
\end{equation}
Equation \eqref{Horizon} assumes non-rotating black holes and takes into account only radiation emitted during the inspiral phase, and assumes that the binary is optimally oriented.  Figure 1 from \citet{LIGO16} shows a plot of the amplitude spectral density (the square root of the power spectral density) of the two LIGO detectors around the time of the GW150914 detection.  We use a table of values estimated from that curve to evaluate the integral in Equation \eqref{Horizon}.  
\par
Using Equation \eqref{Horizon}, we estimate a horizon distance of $D_H = 1.06$ Gpc for GW150914.  This distance is reasonably compatible with the expected result of 1.23 Gpc based on the claim in \citet{Abbott16a} that the merger was detected with SNR 24 and an estimated distance of 410 Mpc, and the expectation that the signal falls off inversely with distance. 
\par
 \begin{figure}
\centering
\includegraphics[width=.8\textwidth]{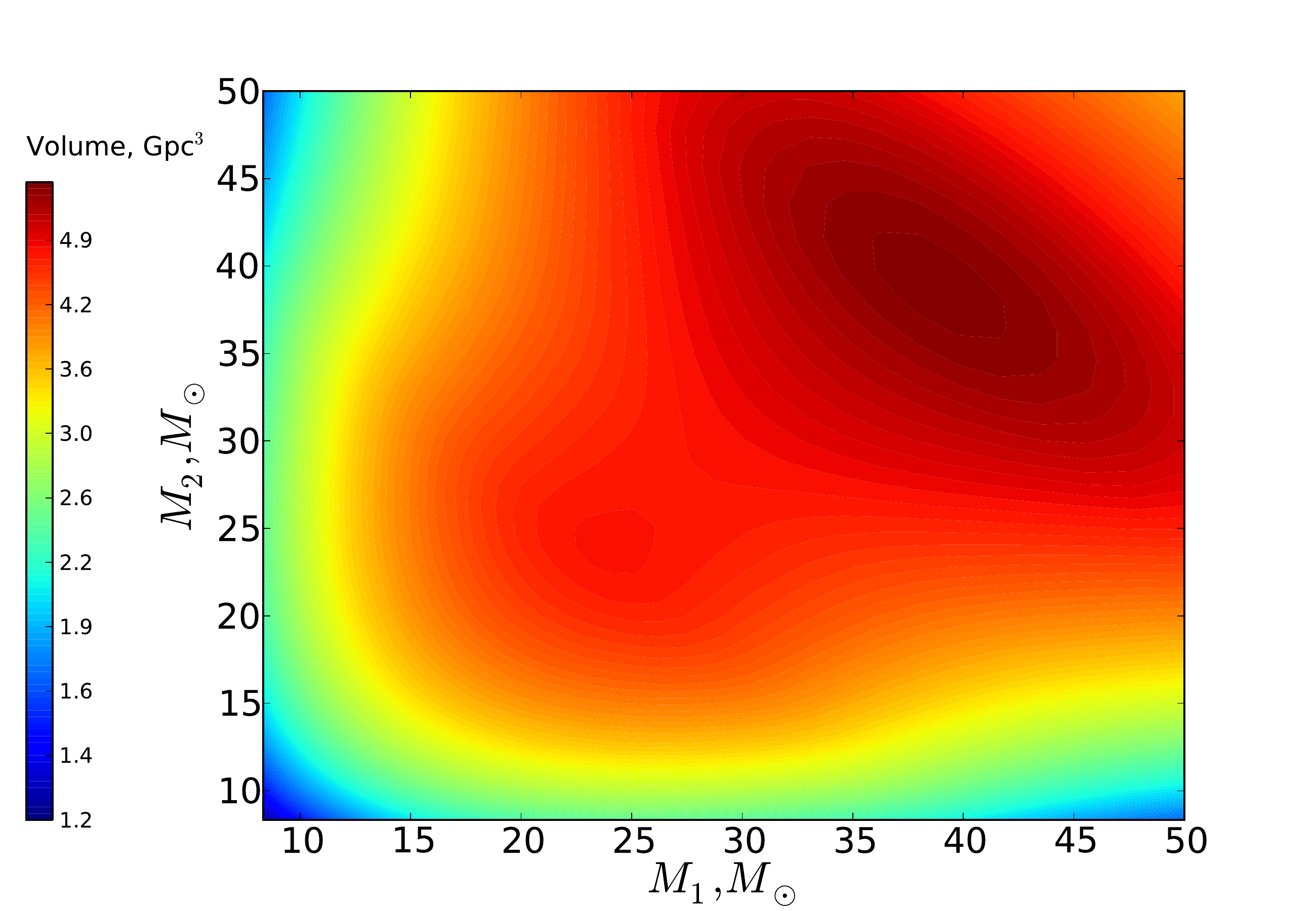}
\caption{Horizon volume as a function of the masses of the merging black holes.  The horizon volume was calculated from the horizon distance obtained using Equation \eqref{Horizon}.}
\label{horizonPlot}
\vspace{-.05cm}
\end{figure} 
To convert from mergers per Gpc$^3$ per year to the expected detection rate $N_{\rm detect}$ given current detectors, we multiply $N_{\rm merge}$ by $\langle V_H\rangle$, where $V_H = 4 \pi D_H^3/3$ is the volume contained within the horizon.  The angle brackets represent an average over all simulated mergers.  For the run with no natal kicks we find $\langle V_H \rangle = 3.2$ Gpc$^3$, giving a value of $N_{\rm detect}$ of 19 per year.  We are ignoring the expansion of the universe and the time evolution of the star formation rate.
\par
Figure \ref{horizonPlot} shows the horizon volume $V_H$ as a function of the masses of the two merging black holes.  The horizon volume is nearly independent of the masses over a wide range because larger black holes, while producing more strain, also emit at lower frequencies to which aLIGO is less sensitive.
\par
Considering just the detection of GW150914, \citet{Abbott16b} give a symmetric 90\% credible interval of 2--53 black-hole binary mergers per comoving Gpc$^3$ per year, assuming all black-hole binaries to have the same masses and spins as the binary that yielded GW150914.  They also analyze the data from many triggers that do not have sufficient S/N to count as detections and estimate a rate of 6--400 black-hole binary mergers per comoving Gpc$^3$ per year.  
\par
We find therefore, that our estimated six mergers per year per Gpc$^3$ due to dynamical evolution in isolated triple systems can account for a substantial fraction, or even all, of the total black-hole mergers, depending on where the true rate lies in the estimated range.  Furthermore, the channel discussed in this paper may be the dominant source of high-eccentricity mergers (mergers that are detected with orbital eccentricity different from zero) over the channels discussed in the literature.  These mergers are discussed further in Section \ref{eccentricMerge}.

\FloatBarrier
\subsection{Eccentricity in the LIGO frequency band}
\label{eccentricMerge}
 \begin{figure}
\centering
\includegraphics[width=.8\textwidth]{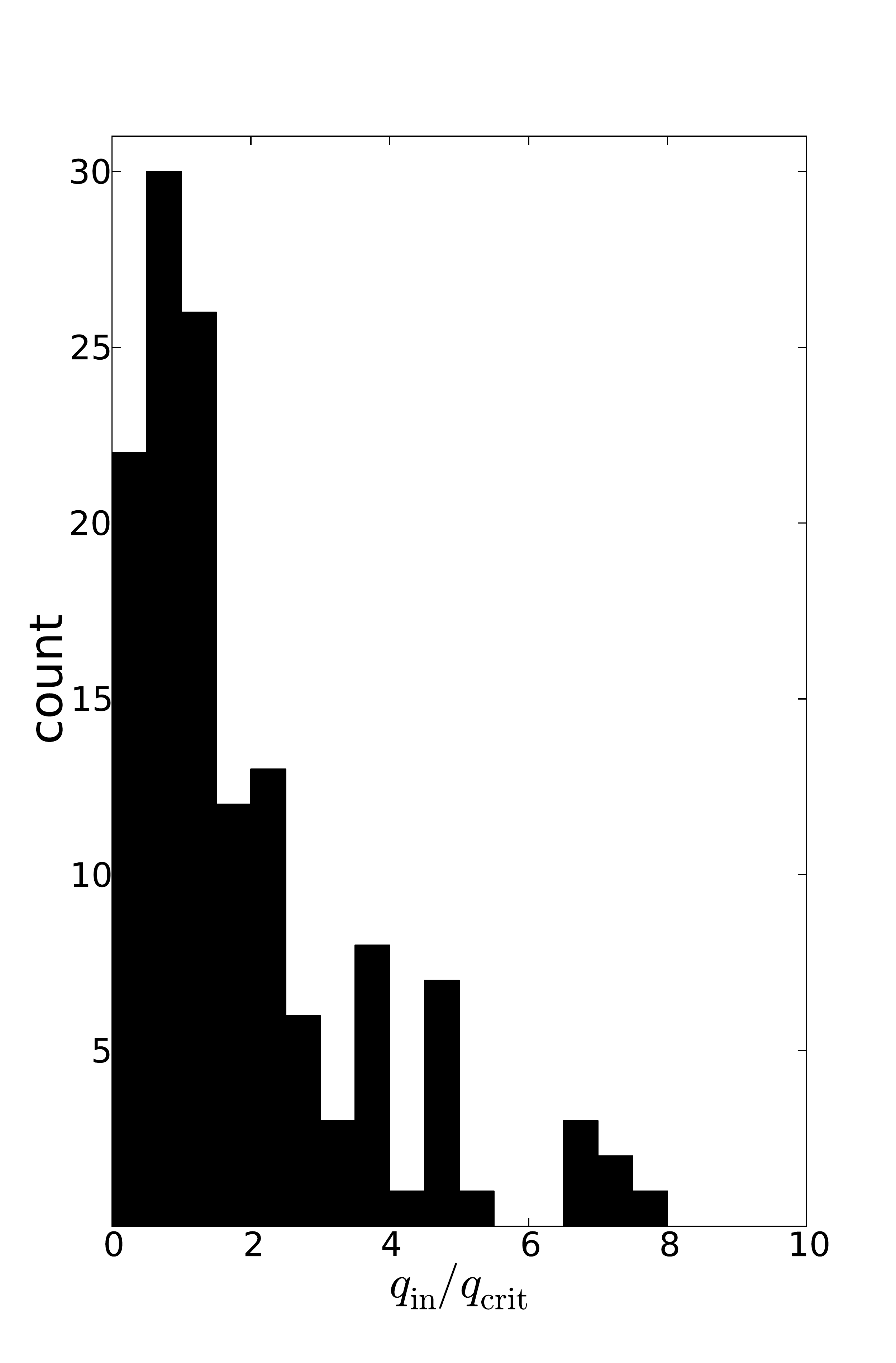}
\caption{Distribution of pericenter distance $q_{\rm in}$ during the initial phase of inspiral, normalized by the rough expected maximum value given in Equation \eqref{qcrit}.  These data represent 135 mergers out of 5,000 total systems simulated with zero kick velocity.  There is one additional system with $q_{\rm in}/q_{\rm crit} = 13.1$.}
\label{RratioHist}
\vspace{-.05cm}
\end{figure}
There is a portion of the inspiral that occurs at nearly constant pericenter distance (see Equation (11) of \citet{Peters64}).  This occurs for values of the semi-major axis such that the apocenter distance has shrunk enough that the outer companion exerts negligible torque on the inner, but the inner orbit is still sufficiently eccentric that the gravitational-wave emission has little effect on the pericenter distance (see Figure \ref{Inspiral2}).  Figure \ref{RratioHist} shows the distribution of the ratio of the pericenter distance of the inner orbit during the phase of constant pericenter distance to the maximum distance given by Equation \eqref{qcrit}.  We see that the 90/136 of cases agree with Equation \eqref{qcrit} to within a factor of two, which is about as good as would be expected. 
\par
We can relate the pericenter distance of the inner orbit to the frequency of emitted radiation with the most power per unit frequency --- the peak frequency.
\citet{Wen03} gives an expression for the peak frequency of emission from an eccentric orbit:
\begin{equation}
\label{freq}
f_{\rm GW}^{m}(e_{\rm in}) = \frac{\sqrt{GM_{\rm in}}}{\pi} (1+e_{\rm in})^{1.1954} \frac{1}{(a_{\rm in}\epsilon)^{1.5}},
\end{equation}
where $\epsilon = 1-e_{\rm in}^2$.  Equation \eqref{freq} can be re-written as 
\begin{equation}
\label{freqinq}
f_{\rm GW}^{m} = \frac{2\nu_{\rm circ}(q_{\rm in})}{(1+e_{\rm in})^{0.3046}},
\end{equation}
where $\nu_{\rm circ}(q_{\rm in}) = \sqrt{GM_{\rm in}}/(2\pi q_{\rm in}^{3/2})$ is the inverse orbital period of a circular orbit with orbital radius equal to the pericenter distance $q$.
Advanced LIGO is sensitive to frequencies as low as 10 Hz.  Equation \eqref{freqinq} implies that a binary with pericenter distance less than 
\begin{equation}
\label{qmin}
q_{\rm min} = 1522 \left(\frac{M_{\rm in}}{40 M_\odot}\right)^{1/3} {\rm km}
\end{equation}
will emit in the LIGO band regardless of its eccentricity.  We define a ``high-eccentricity" merger as a merger where, at the time the simulation terminates (see Section \ref{termination}), the pericenter distance calculated from Equation \eqref{qofL} is less than $q_{\rm min}$ defined in Equation \eqref{qmin}.  Using this criterion, we found that 2/136 of the merger events in our simulations with no natal kicks, or 12/416 in all our simulations were ``high-eccentricity" mergers.  One might expect this fraction to be equal to $\langle q_{\rm min}/q_{\rm crit}\rangle$, where the angle brackets represent an average over all merging systems.  However, the fraction of merger events seen in simulation is smaller than would be predicted from Equations \eqref{qcrit} and \eqref{qmin} by a factor of 4.6 for the set of simulations with no natal kicks, and by a factor of 2.1 for all simulations. 
\par 
 One explanation for this discrepancy could be that our approximation that $\Delta L_{\rm in}/\Delta L \gg 1$ is not valid.  In the opposite limit of $\Delta L_{\rm in}/ \Delta L  \ll 1$ one would expect no high-eccentricity mergers.  We also see in Figure \ref{RratioHist} that Equation \eqref{qcrit} seems to underestimate the critical pericenter distance leading to merger, which leads to an overestimate of the number of expected high-eccentricity mergers.  It is worth noting that it has been found \citep{Seto13, Antognini14, Antonini14} that mergers where the inner binary has high eccentricity when radiating at frequencies above 10 Hz are only found when the orbits are simulated with an N-body code, rather than an octupole secular code.
\FloatBarrier
\subsection{Delay time distribution}
\label{sect:dtd}
 \begin{figure}
\centering
\includegraphics[width=.8\textwidth]{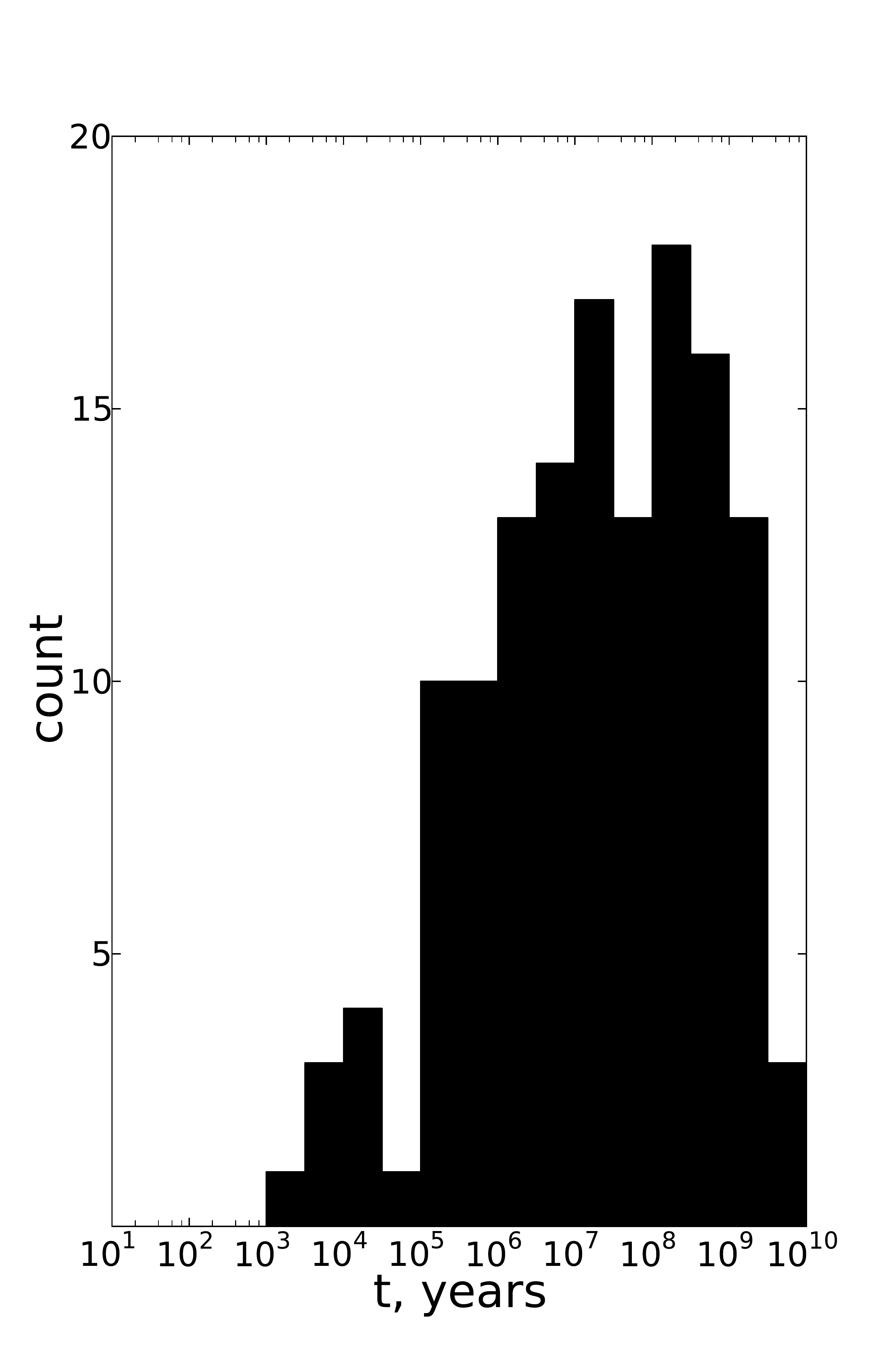}
\caption{Distribution of delay time - the time between the initialization of the simulation and the merger.  This shows the results for 136 mergers, with no natal kicks.}
\label{absoluteTime0}
\vspace{-.05cm}
\end{figure} 

 \begin{figure}
\centering
\includegraphics[width=.8\textwidth]{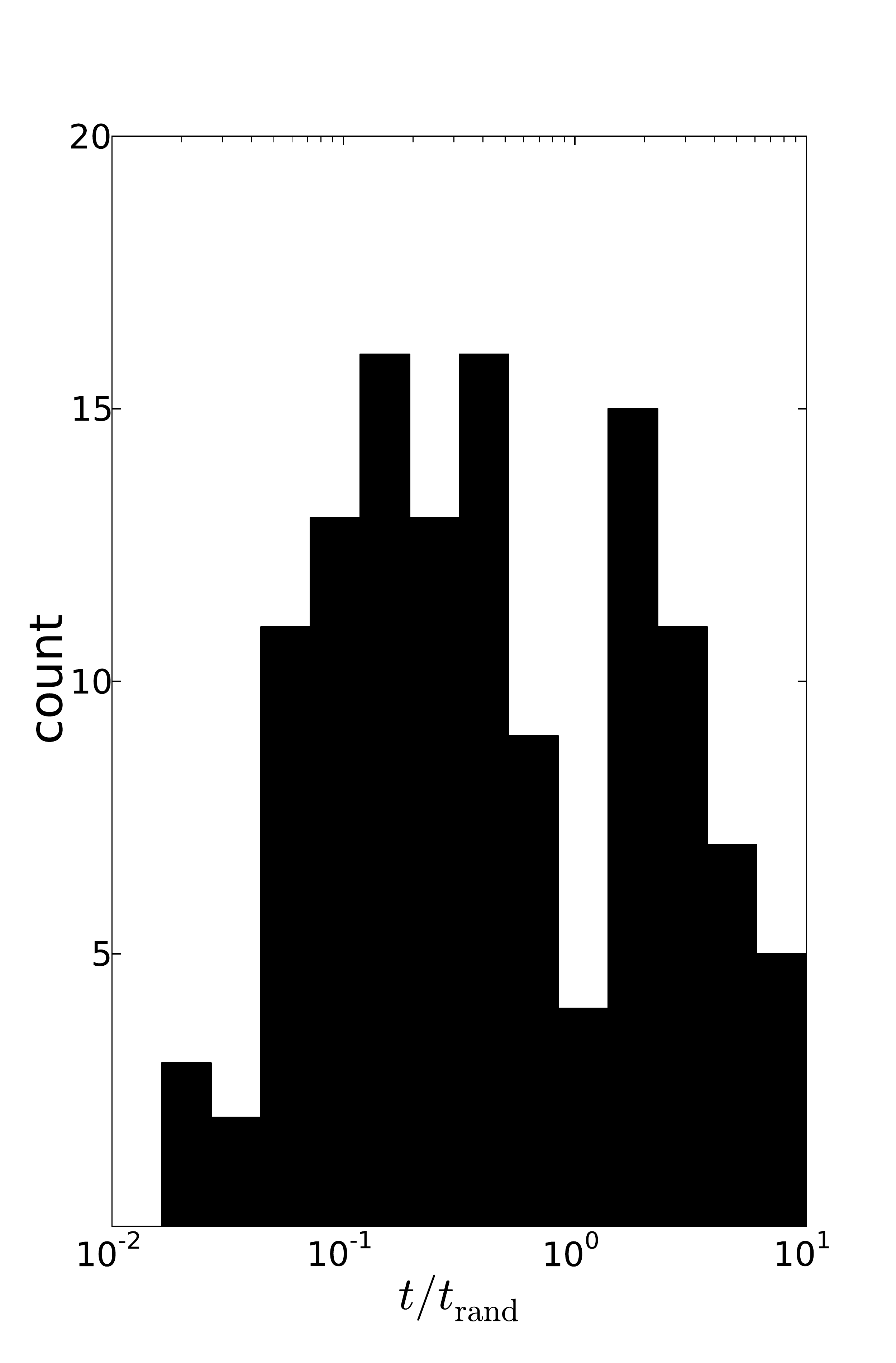}
\caption{Distribution of delay time normalized by $t_{\rm rand}$, defined in Equation \eqref{trand}.  This shows the results for 125 mergers, with no natal kicks.  11 additional systems merged in less than $0.01 t_{\rm rand}$.}
\label{timehist}
\vspace{-.05cm}
\end{figure}

 \begin{figure}
\centering
\includegraphics[width=.8\textwidth]{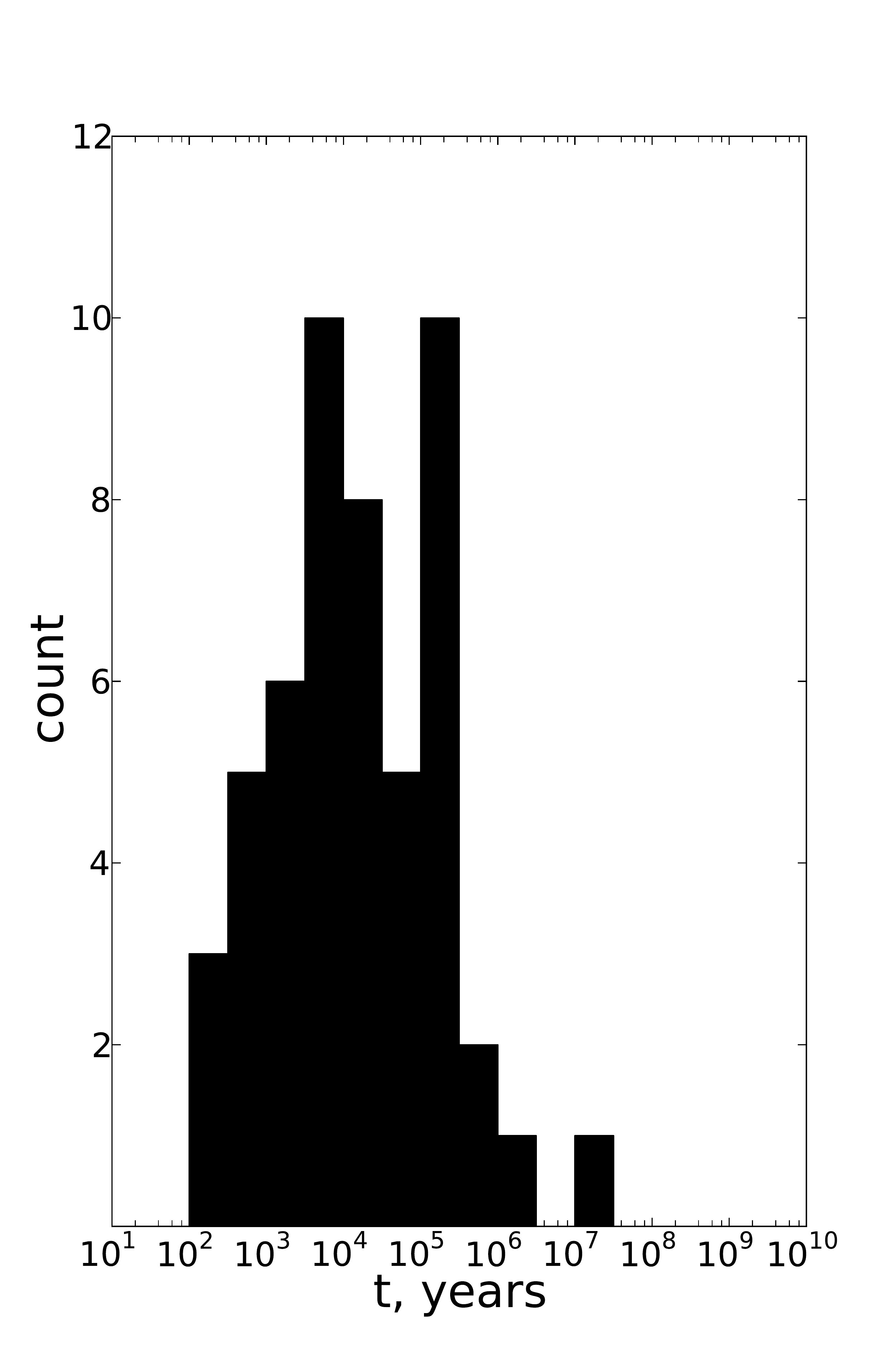}
\caption{Distribution of delay time - the time between the initialization of the simulation and the merger.  This shows the results for 51 mergers, with natal kicks of 40 km s$^{-1}$.  Note that the distribution has shifted to smaller times relative to Figure \ref{absoluteTime0}, because less compact systems (which take longer to merge) are preferentially unbound by the natal kick.}
\label{absoluteTime40}
\vspace{-.05cm}
\end{figure}

Assuming that most mergers occur in a time similar to $t_{\rm rand}$ (see Equation \eqref{trand}), and assuming furthermore that $\epsilon_{\rm merge}$ is independent of $a_{\rm in}$, we can use Equations \eqref{qcrit} and \eqref{trand} to estimate the distribution of delay times.  Under these assumptions, we would expect the delay time distribution to be roughly flat in log $t_{\rm delay}$ with $t_{\rm delay}$ running over the range of possible values of $t_{\rm rand}$.  The value of $t_{\rm rand}$ is determined mostly by $a_{\rm in}$, which ranges from about 4 to 4,000 AU for the systems we simulate.  For a system of three 20 $M_\odot$ black holes with $a_{\rm out} = 5 a_{\rm in}$, $a_{\rm in} = 4$ AU corresponds to $t_{\rm rand} = 3.5 \cdot 10^4$ years, and $a_{\rm in} = 4,\!000$ AU corresponds to $t_{\rm rand} = 2.0\cdot 10^{11}$ years.  These arguments suggest that systems with values of $a_{\rm in} $ larger than 4,000 AU would rarely merge within the age of the universe even if they did exist.
\par
 Figure \ref{absoluteTime0} shows the delay time distribution of our merger events.  We see that as suggested in the previous paragraph, the distribution of delay times is roughly flat in $\log(t_{\rm delay})$ above $t_{\rm delay} = 3 \cdot 10^4$ years but there are some outliers at very short delay times.  The majority of events occurring at the present time are due to relatively recent star formation (within the last Gyr, corresponding to $z \leq 0.075$), which confirms our assumption that we can calculate the rate using the local star formation rate, as we did in Equation \eqref{rateEquation}.  
\par
Figure \ref{timehist} shows the distribution of delay time normalized by $t_{\rm rand}$.  This gives some indication of the way in which the inner orbit explores phase space.  In a model in which the space of squared eccentricity is explored randomly and uniformly, we would expect the delay time distribution to be exponential with mean $t_{\rm rand}$.  What we see is that more than half of systems that merge do so in the first interval $t_{\rm rand}/3$, yet 8/136 systems take between $5t_{\rm rand}$ and $10t_{\rm rand}$ to merge.  These facts are inconsistent with a model in which eccentricity space is explored randomly and uniformly.  This discrepancy is not surprising, given that we know all of phase space is not explored for all systems, since the vast majority of them do not merge at all.  The existence of systems that do not merge until several $t_{\rm rand}$ shows that the correction to the simple model is more complicated than just a restriction on the accessible phase space. 
\par
 The distribution shown in Figure \ref{timehist} raises the question of whether we have run the simulation long enough to see all the mergers that would occur.  We believe that we have missed fewer than 1/3 of the mergers, assuming a constant number of mergers per log of $t_{\rm sim}/t_{\rm rand}$ beyond $10t_{\rm rand}$, and assuming that we would naturally not be interested in mergers that take longer than $10^{10}$ years.  It is much more probable that the number of mergers per log $t_{\rm sim}/t_{\rm rand}$ continues to decline beyond $t_{\rm sim}/t_{\rm rand} = 10$, and our underestimate is less serious.
  \par
It is worth treating with suspicion the systems that merge within $3 \cdot 10^6$ years (about the main-sequence lifetime of a massive star), as one must answer the question of why such a system did not merge on the main sequence, as in e.g., \citet{Naoz15}.  It is certainly possible that the orbital elements were changed during the late stages of stellar evolution, or by mild natal kicks.  We note that removing from consideration all mergers that occur within $3 \cdot 10^6$ years cuts our rate by 30\%. 
\par
 In actuality, it is possible that a fraction significantly higher than 30\% merges while the black holes are still stars - not only are the physical radii of the stars substantially larger than $q_{\rm crit}$, but also the orbits expand, and the stars become less massive in our model as a result of mass loss at the end of stellar evolution.  
\par
Including natal kicks shortens the delay time distribution by preferentially unbinding more widely separated systems.  This is shown in Figure \ref{absoluteTime40}, where we see that almost all mergers occur within $10^7$ years if we assume natal kicks with rms velocity 40 km s$^{-1}$.  Natal kicks also reduce the probability that a system which merges quickly would have merged on the main sequence, as the orbital elements are altered by the kick.

 \FloatBarrier
 \subsection{Inclination distribution}

 \begin{figure}
\centering
\includegraphics[width=.8\textwidth]{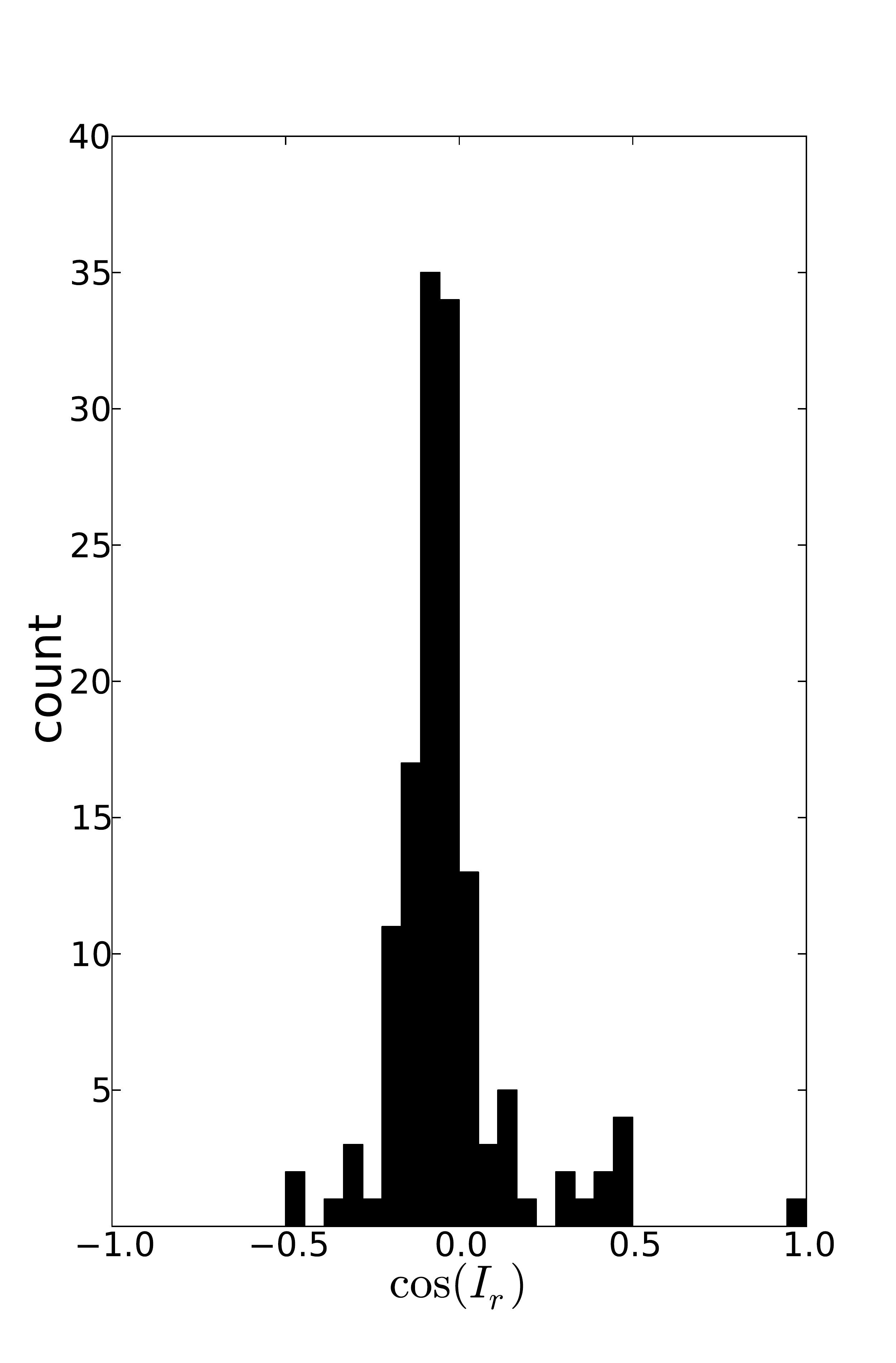}
\caption{Distribution of the cosine of the initial mutual inclination between the inner and outer orbits.  These data represent 136 merging systems out of 5,000 simulated systems, with zero natal kick.}
\label{IncHist}
\vspace{-.05cm}
\end{figure} 
Figure \ref{IncHist} shows the distribution of initial relative inclination between the inner and outer systems for the subset of systems which do merge.  In the quadrupole approximation to Lidov-Kozai oscillations, mergers would only occur if the initial relative inclination between the inner and the outer orbit were within $\approx \sqrt{2 q_{\rm crit}/a_{\rm in}}$ of 90$^\circ$ (corresponding to $0.14^\circ$ for the system in the numerical estimate of Equation \eqref{qcrit} with $a_{\rm in} = 30$ AU).  In actuality however, we see that while there is a strong preference for initially highly inclined systems, mergers occur with some probability for systems with a wide range of initial inclinations.

\subsection{Spin alignment}
\citet{Kushnir16} are able to rule out some formation channels for GW150914 based on the small value of the mass weighted average spin vector projected on the orbital angular momentum.  It is therefore worth considering what spin-orbit alignment properties binaries merging due to the channel discussed in this paper would be expected to exhibit. 
\par
  \citet{Hale94} states that solar type stars in binaries with separations under about 30 AU generally have spins aligned with the orbit to within 10 degrees.  23\% of our merging systems have initial separations less than 30 AU (or about 40\% if we assume a critical separation corresponding to the period of a solar mass binary with separation 30 AU).  However, we find that torques from the tertiary companion nearly isotropize the binary orbital plane relative to its original value, thus eliminating the alignment between the orbit and the spins.
    \par
    The spins will also be tilted relative to one another due to geodetic precession.  \citet{Barker75} give the orbit-averaged rate of precession of the spin of one component of the binary around the orbital angular momentum axis:
\begin{equation}
\label{desittereq}
\dot \Omega_{ds} = \frac{3Gn(m_j+ \mu/3)}{2c^2a(1-e^2)}.
\end{equation}
Here, $n$ is the mean motion of the binary, and $m_j$ is the mass of the companion.
\par
We can use Equation \eqref{desittereq} in conjunction with Equation \eqref{tauMerge} to estimate the amount of precession that occurs during a circular inspiral from initial separation $a$:
\begin{equation}
\Delta \Omega_{ds} \approx \tau_{\rm merge} \dot \Omega_{ds} =  \frac{15}{512} \left(\frac{m_j}{\mu} + 1/3 \right) \left(\frac{c}{v_{\rm circ}(a)}\right)^{3},
\end{equation}
where $v_{\rm circ}(a) = \sqrt{GM/a^3}$.  Plugging in fiducial parameters ($m_1 = m_2 = 20 M_\odot$), we find that $\Delta \Omega_{ds} = 217$, assuming a circular inspiral starting from our fiducial value of $a = q_{\rm crit} = 12,\!800$ km.  If the two bodies were exactly the same mass, then their spins would precess by the same amount and remain aligned, but even a few percent deviation in mass is sufficient to cause a $180^\circ$ misalignment.  The majority of the precession occurs when $a$ is near its initial value, therefore before the system would be detected by LIGO.  We therefore expect the spins to show little correlation with each other, as well as little correlation with the orbital plane.
\par
\begin{figure}
\centering
\includegraphics[width=.8\textwidth]{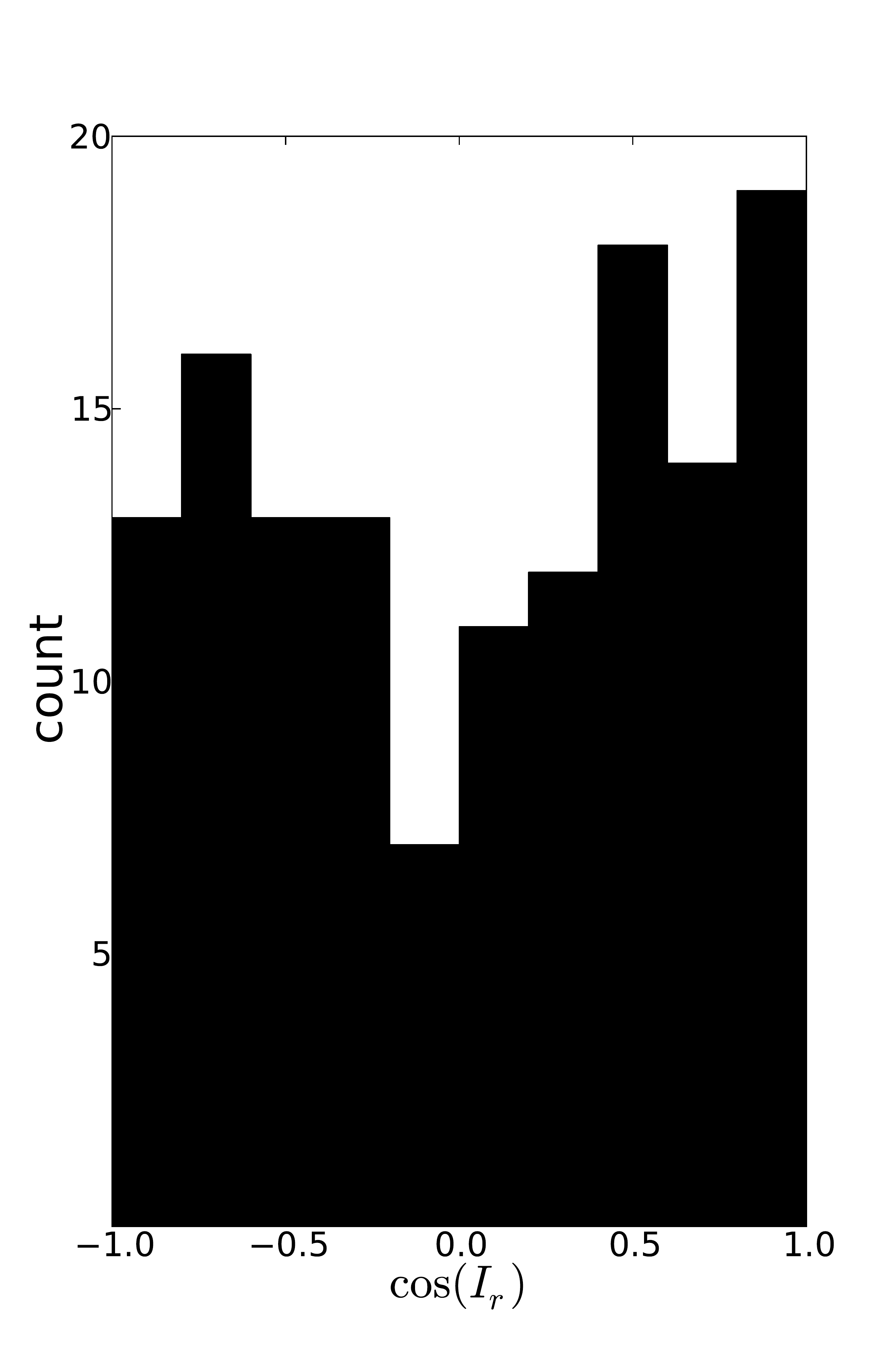}
\caption{Distribution of the relative inclination between the inner orbit at the beginning of the numerical integration and the inner orbit at the time of merger.  These data represent 136 merging systems out of 5,000 simulated systems, with zero natal kick.}
\label{alignmentHist}
\vspace{-.05cm}
\end{figure}

 Figure \ref{alignmentHist} shows the cosine of the inclination of the inner orbit at the time of merger, relative to the plane of the inner orbit at the beginning of the simulation.  We see that the relative inclination distribution is nearly isotropic.  

\FloatBarrier
\subsection{Distribution of the stability ratio $R_s$ among merging systems}
 \begin{figure}
\centering
\includegraphics[width=.8\textwidth]{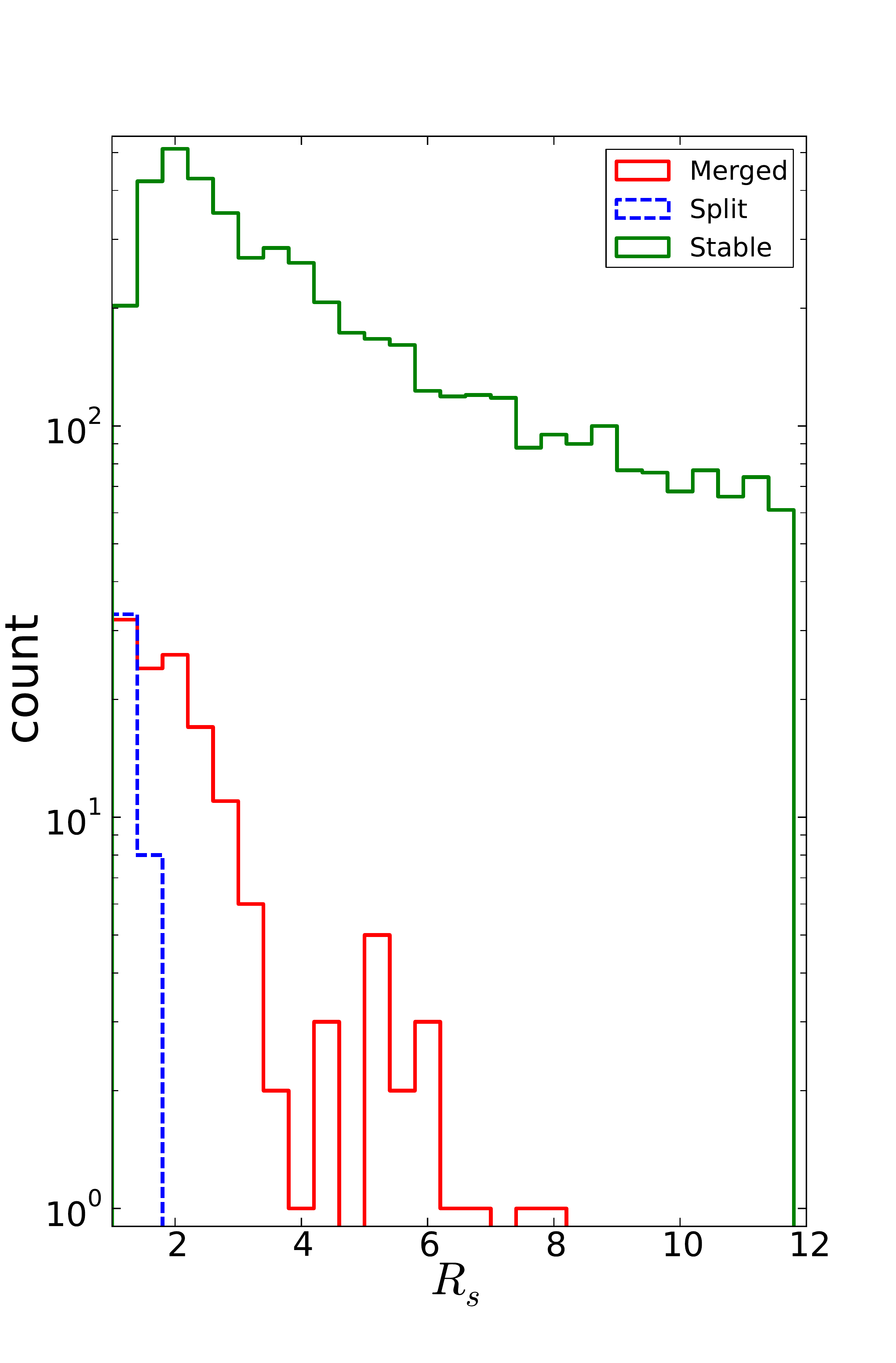}
\caption{Distribution of fates of the systems (merged, ejected one member, or stable against merging and ejection) as a function of the initial stability ratio defined in Equation \eqref{Rs}.  This figure was made for 5,000 systems, assuming no natal kick.  There are fewer total systems close to the stability boundary $R_s = 1$, since some are eliminated during the late stages of stellar evolution if the inner orbit temporarily expands by a larger factor than the outer.}
\label{outcomes}
\vspace{-.05cm}
\end{figure}
At the end of the simulation period, we classify systems as either ``merged" (two of the stars merged together in the simulation), ``split" (one or more of the stars moved outside the simulation region), or ``stable" (neither of the two above outcomes occurred). Figure \ref{outcomes} shows the distribution of outcomes as a function of the initial stability ratio $R_s$, as defined in Equation \eqref{Rs}.  We see that the majority (110/136) of merger events occur for $R_s <3$, but there is a substantial tail of events occurring at larger values of $R_s$.  In our simulations with no kick velocity, we see no merger events with $R_s > 8$, so we believe that we are not missing a significant fraction of events by not considering systems with $R_s > 12$.  Since the stability criterion in Equation \eqref{eq:Ycrit} is not perfect, it is likely that we are further underestimating our rates by not simulating any systems with $R_s < 1$.
\FloatBarrier
\subsection{Chirp mass and mass ratio of merging systems}
\label{chirpAndRatio}
We additionally ask what masses of black holes are most likely to merge.  We characterize the masses of the merging binary black hole by the mass ratio $\kappa_{\rm in}$, and the chirp mass given in Equation \eqref{Mchirp}. 
\par
The number of merging systems seen in our simulations as a function of $\kappa_{\rm in}$ and $M_{\rm chirp}$ is strongly dependent on assumptions about statistics of triple systems that are not well constrained by observations.  For this reason, we choose to present results normalized to the number of such systems that we simulated, so we can see the relative merger rates.
\par
\begin{figure}
\centering
\includegraphics[width=.8\textwidth]{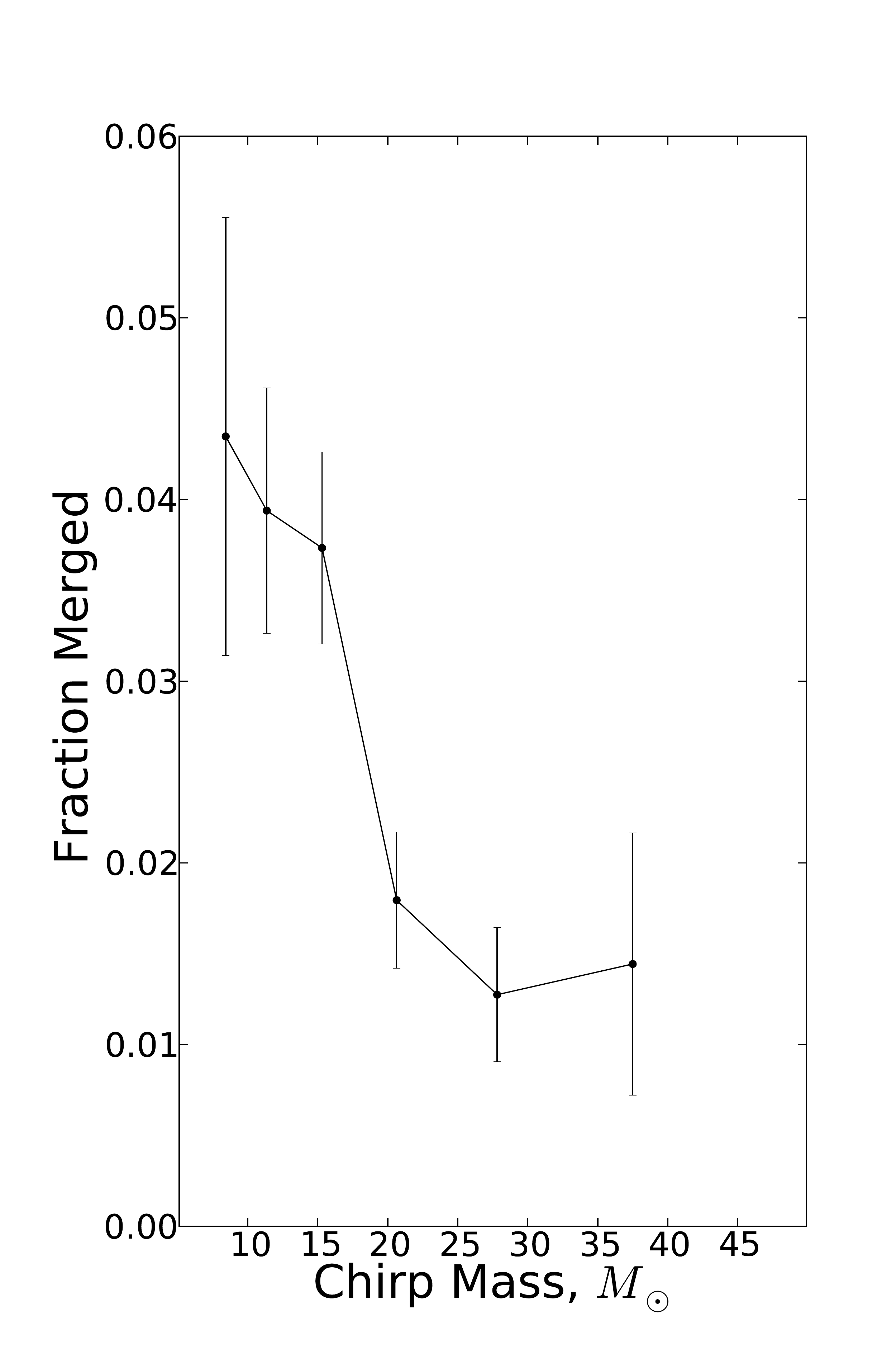}
\caption{Fraction of simulated systems that merge as a function of chirp mass.  These data were taken from 136 mergers out of 5,000 total systems, assuming no natal kicks.}
\label{chirpPlot0}
\vspace{-.05cm}
\end{figure}

Figure \ref{chirpPlot0} shows $\epsilon_{\rm merge}$: the fraction of simulated systems that actually merge, as a function of the chirp mass.  The error bars are equal to $\epsilon_{\rm merge}/\sqrt{N_{\rm merge}}$, where $N_{\rm merge}$ is the number of merger events in a given bin.  This assumes that the number of mergers in a sample run is given by a Poisson distribution, i.e., that $\epsilon_{\rm merge} \ll 1$.  The exception to this rule is for bins in which $N_{\rm merge}$ is 0.  In this case, the error bar is reported as $1/N_{\rm tot}$, where $N_{\rm tot}$ is the total number of systems in that bin.  If $\epsilon_{\rm merge} > 1/N_{\rm tot}$, then it requires an unlikely event to see no mergers in that bin.
\par
We see that in general, systems with larger chirp mass are less likely to merge, though the dependence is not strong.  Systems with a large chirp mass generally have a higher fraction of the angular momentum in the inner binary, and thus require more fine-tuned initial conditions to yield a merger.  

 \begin{figure}
\centering
\includegraphics[width=.8\textwidth]{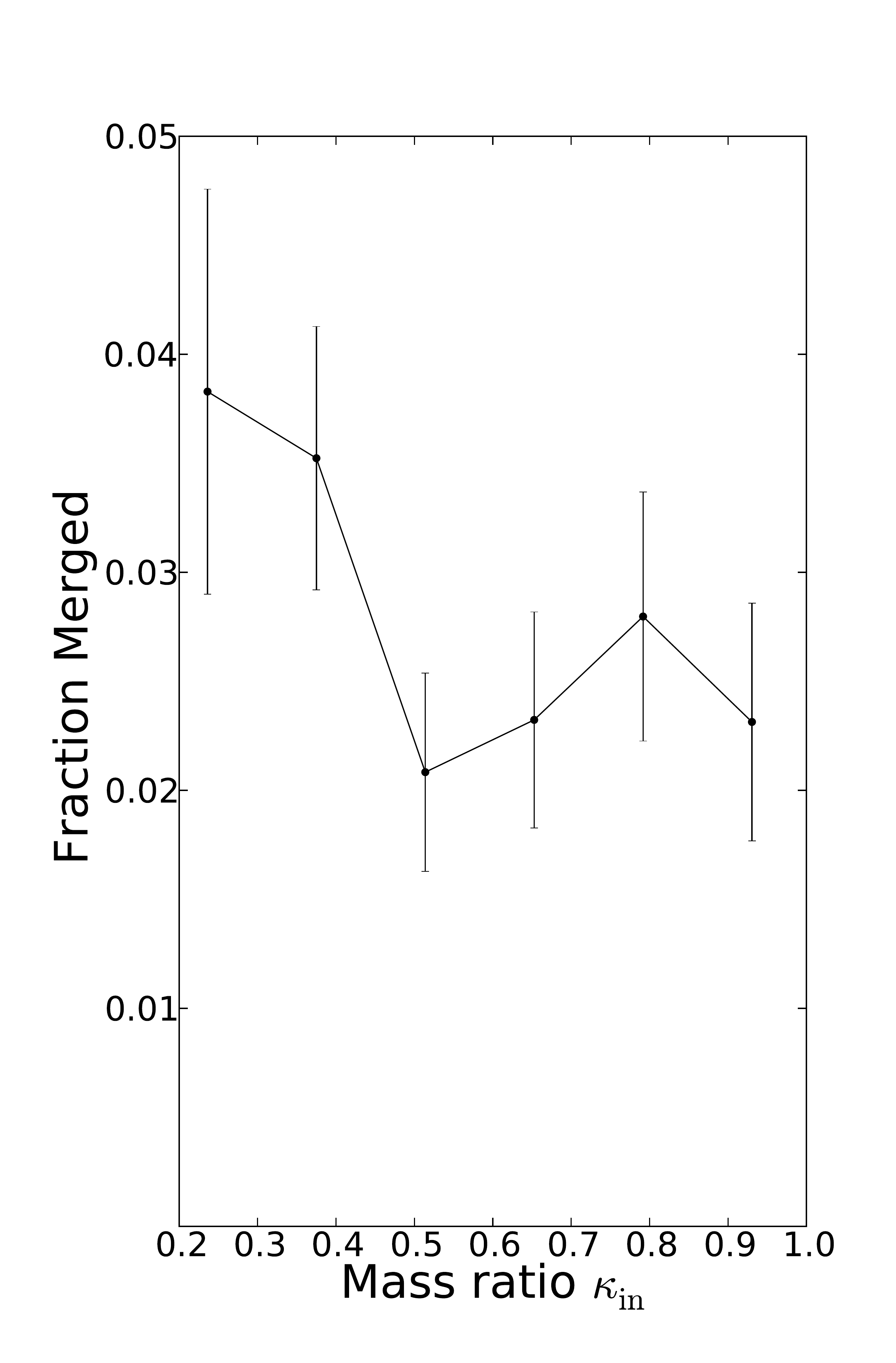}
\caption{Merger rate as a function of mass ratio.  These data were taken from 136 mergers out of 5,000 total systems, assuming no natal kicks.}
\label{qPlot0}
\vspace{-.05cm}
\end{figure}
Figure \ref{qPlot0} shows $\epsilon_{\rm merge}$ as a function of the mass ratio $\kappa_{\rm in}$.  The error bars are calculated in the same way as for Figure \ref{chirpPlot0}.  Systems with very small mass ratio inner binaries have a marginally higher merger fraction, probably because they have less angular momentum in the inner orbit relative to the outer orbit for a given semi-major axis ratio.

\section{Other models for the pre-black-hole evolution}
\label{otherModels}
\FloatBarrier
\subsection{Natal kicks}
\label{kicks}
In the previous sections we have considered systems in which the black holes receive negligible velocities during their formation, and in which mass loss is assumed to take place on a timescale long compared with the orbits of the inner and outer binaries.  It is possible however, that a black hole receives a kick due to recoil from an asymmetric supernova explosion.  Even in a symmetric supernova, a binary receives a kick to its center of mass in a supernova explosion because one of the massive moving components suddenly loses mass.  This mechanism, called a ``Blaauw" kick \citep{Blaauw61}, does not operate in our simulations because we assume that mass loss occurs adiabatically prior to black hole formation.  Since the typical velocity of the outer orbit is kilometers per second to tens of kilometers per second, we expect kick velocities of these magnitudes to unbind the systems, thus sharply reducing the estimated rate of mergers.   There has been some work estimating these velocities, but the results are far from conclusive.
\par
From the position and velocity of an X-ray binary, its path can be traced back to the Galactic plane, and (assuming it was born in the plane) the natal kick can be estimated.  \citet{Brandt95b} study the case of the soft X-ray transient GRO J1655---40 and conclude that the system likely received a kick of about 100 km s$^{-1}$ soon after its formation.  \citet{Williams05} do additional modeling and conclude that the combination of a Blaauw kick and a possible natal kick gave the center of mass of the system a kick between 45 and 115 km s$^{-1}$.  They conclude that the natal kick was between 0 and 210 km s$^{-1}$. 
\par
\citet{Repetto15} perform a similar analysis on seven short period X-ray binaries, and conclude that in two of them either a high natal kick ( $\sim 100$ km s$^{-1}$ and $\sim 400$ km s$^{-1}$) or ejection from a globular cluster (as opposed to being born in the plane of the Galaxy) is required, whereas the others could be consistent with no natal kick.  \citet{Mandel16a} questions the assumptions in \citet{Repetto15} and claims that the data in \citet{Repetto15} do not require kicks in excess of 80 km s$^{-1}$ although higher kicks are not ruled out.
\par
 It is thought that at least many neutron stars receive large ($> 100$ km s$^{-1}$) natal kicks e.g., \citep{Brandt95a,  Willems04}.  However, there is evidence for a population that do not.  \citet{Beniamini16} analyze the orbits of 10 Galactic double neutron star binaries, and conclude that there is evidence for a substantial population of neutron stars that receive kicks on the order of 5 km s$^{-1}$.
\par
We have performed simulations using the procedure outlined in Section \ref{massLoss} for different values of the characteristic kick velocity $v_{\rm rms}$.  In our simulations, we find that of the systems that survive the kicks, the merger rate is roughly independent of the kick velocity, but kick velocities greater than $\sim 10$ km s$^{-1}$ unbind most of the systems under consideration.  The results are presented in Table \ref{table:yields}.
\par
Our estimated rates are a factor of about 10 higher than the rates estimated for the scenario in which triples form in the center of globular clusters.  The existence of natal kicks would tend to lower the rates there as well.  We study the list of 141 Galactic globular clusters provided at \url{http://dept.astro.lsa.umich.edu/~ognedin/gc/vesc.dat}, associated with \citet{Gnedin02}.  We find that the mass-weighted median central escape velocity from these clusters is 48 km s$^{-1}$.  This speed is certainly higher than the orbital speeds of our triple systems, but only by factors of a few, implying that if natal kicks would prevent formation of merging black-hole binaries in triple systems, they are also likely to do so in globular clusters.
\FloatBarrier
\subsection{Other models for picking masses and semi-major axes}
\label{otherModels}
Denote the model for picking masses and semi-major axes described in Section \ref{setup} as model A.  We also consider a more standard model in which for each triple, the mass $m_1$ of one of the stars in the inner binary is picked from the IMF.  Then $\kappa_{\rm in}$ and $\kappa_{\rm out}$ are drawn from the distribution in Equation \eqref{Nqm}.  Then $m_2 = \kappa_{\rm in} m_1$ and $m_3 = \kappa_{\rm out} m_1$.  The standard method of picking the masses lowers the rates by about a factor of 1.5 compared with the method described in Section \ref{masses}.  This is at least in part because stars placed in multiple systems according to the standard method are preferentially smaller than stars drawn from the IMF.  Stars in multiple systems picked with our preferred method have the same mass distribution as the IMF.  One might argue that stars in multiple systems should be preferentially {\it more} massive rather than {\it less} massive than single stars given the general trend of massive stars having higher multiplicity.  The detailed results of this selection process are shown in the fifth row of Table \ref{table:yields} corresponding to ``model B".  
\par
 In section \ref{chirpAndRatio} we found that the merger rates are not strongly dependent on the black hole masses.  We therefore believe that the largest effect of changing our model for choosing the masses of the stars, or the initial-final mass function, would arise through changes in the number of triple black hole progenitor systems per solar mass of new stars.  This ratio is $5.9\cdot 10^{-5}$ and $4.5 \cdot 10^{-5}$ for our models A and B respectively.
\par
We also tested the sensitivity of our results to the upper limit on the separation of the outer orbit.  Model C in Table \ref{table:yields} has the same parameters discussed in Section \ref{setup} except the upper limit on the semi-major axis of the outer orbit is 700 AU instead of 6,000 AU.  We see that this choice makes little difference.

\begin{table}[ht]
\caption{Merger rates for a variety of models.  $v_{\rm rms}$ is the RMS natal kick velocity (see Equation \eqref{sigmakdist}).  $\langle V_H \rangle$ is the expected volume over which a given merger could be detected by aLIGO.  $\epsilon_{\rm suit}$ is the fraction of total triple black-hole systems that we believed might merge, and therefore simulated.  $\epsilon_{\rm merge}$ is the fraction of simulated systems that actually merged.}  
\centering
\vspace{1mm}
\begin{tabular}{c c c c c c c c}
\hline \hline
Model&$v_{\rm rms}$ (km s$^{-1}$)&$\epsilon_{\rm trip}$& $\epsilon_{\rm suit}$&$\epsilon_{\rm merge}$&$V_H$ (Gpc$^3$)&N (yr$^{-1}$ Gpc$^{-3}$)&$N_{\rm detect}$ (yr$^{-1}$) \\ [0.5ex] 
\hline
A & 0   & 0.0258 & 0.151     & 0.0272 & 3.20 & 6.1   & 19.4  \\
A & 10 & 0.0258 & 0.0388   & 0.0315 & 3.69 &1.8    & 6.6\\
A & 20 & 0.0258 & 0.0135   & 0.028 & 3.93 & 0.56 & 2.2 \\
A & 40 & 0.0258 & 0.00362 & 0.0255 &4.46  & 0.14 & 0.61\\ 
B & 0   & 0.0198 & 0.141      & 0.0245 &3.38  & 3.9   & 13.2\\ 
C & 0   & 0.0258 & 0.141      & 0.0305 &3.59  & 6.3   & 22.8\\ [1ex]
\hline
\label{table:yields}
\end{tabular}
\end{table}
\section{Conclusions}
  We estimated the rate of black-hole mergers in isolated triple systems in the field to be 6 per Gpc$^3$ per year in our fiducial model.  We found that a few percent of moderately hierarchical triple systems yield a merger within a Hubble time for a wide range of black hole masses, however there is much uncertainty about the abundance and orbital parameters of triple systems, as well as the black-hole formation process.  The merger rate depends strongly on the assumed natal kick velocities, declining from 6 per year per Gpc$^{-3}$ assuming no natal kicks, down to 0.14 per year per Gpc$^{-3}$ for 40 km s$^{-1}$ natal kicks, or from 19 down to 0.6 detections per year with the current aLIGO detectors.  We tried a few different models for choosing the masses and semi-major axes of stars in the triple systems and did not find major differences in the merger rate.  However, we have by no means considered the full range of reasonable models given the current data.

\par

The estimated rate without natal kicks is within the (wide) range estimated from the current advanced LIGO results, and about a factor of 10 larger than the rate estimated for formation via the Lidov-Kozai mechanism in globular clusters.  A few percent of these mergers are expected to be highly eccentric, entering the 10 Hz window detectable by LIGO when their eccentricities are $>0.999$.  If natal kicks are small, the mechanism discussed in this paper is likely the dominant source of high-eccentricity mergers among the known channels for producing such events, and could be the dominant source of all binary black-hole mergers.
\par
We thank Todd Thompson and Doron Kushnir for discussions and for comments on the manuscript.  This research was supported in part by NASA grant NNX14AM24G and NSF grant AST-1406166
\bibliographystyle{apj}
\bibliography{apj-jour,kozaiMerger.bib}

\begin{thebibliography}{}
\expandafter\ifx\csname natexlab\endcsname\relax\def\natexlab#1{#1}\fi

\bibitem[{{Abadie, J. {\it et. al.} ({LIGO Scientific Collaboration and Virgo
  Collaboration}})(2012)}]{LIGO12}
{Abadie, J. {\it et. al.} ({LIGO Scientific Collaboration and Virgo
  Collaboration}}). 2012, ArXiv e-prints, arXiv:1203.2674

\bibitem[{{Abbott} {et~al.}(2016{\natexlab{a}}){Abbott}, {Abbott}, {Abbott},
  {Abernathy}, {Acernese}, {Ackley}, {Adamo}, {Adams}, {Adams}, {Addesso}, \&
  et~al.}]{LIGO16}
{Abbott}, B.~P., {Abbott}, R., {Abbott}, T.~D., {et~al.} 2016{\natexlab{a}},
  Classical and Quantum Gravity, 33, 134001

\bibitem[{{Abbott} {et~al.}(2016{\natexlab{b}}){Abbott}, {Abbott}, {Abbott},
  {Abernathy}, {Acernese}, {Ackley}, {Adams}, {Adams}, {Addesso}, {Adhikari},
  \& et~al.}]{Abbott16a}
---. 2016{\natexlab{b}}, Physical Review Letters, 116, 061102

\bibitem[{{Abbott} {et~al.}(2016{\natexlab{c}}){Abbott}, {Abbott}, {Abbott},
  {Abernathy}, {Acernese}, {Ackley}, {Adams}, {Adams}, {Addesso}, {Adhikari},
  \& et~al.}]{Abbott16b}
---. 2016{\natexlab{c}}, ArXiv e-prints, arXiv:1602.03842

\bibitem[{{Antognini} {et~al.}(2014){Antognini}, {Shappee}, {Thompson}, \&
  {Amaro-Seoane}}]{Antognini14}
{Antognini}, J.~M., {Shappee}, B.~J., {Thompson}, T.~A., \& {Amaro-Seoane}, P.
  2014, \mnras, 439, 1079

\bibitem[{{Antonini} {et~al.}(2016){Antonini}, {Chatterjee}, {Rodriguez},
  {Morscher}, {Pattabiraman}, {Kalogera}, \& {Rasio}}]{Antonini16}
{Antonini}, F., {Chatterjee}, S., {Rodriguez}, C.~L., {et~al.} 2016, \apj, 816,
  65

\bibitem[{{Antonini} {et~al.}(2014){Antonini}, {Murray}, \&
  {Mikkola}}]{Antonini14}
{Antonini}, F., {Murray}, N., \& {Mikkola}, S. 2014, \apj, 781, 45

\bibitem[{{Antonini} \& {Perets}(2012)}]{Antonini12}
{Antonini}, F., \& {Perets}, H.~B. 2012, \apj, 757, 27

\bibitem[{{Barker} \& {O'Connell}(1975)}]{Barker75}
{Barker}, B.~M., \& {O'Connell}, R.~F. 1975, \prd, 12, 329

\bibitem[{{Belczynski} {et~al.}(2002){Belczynski}, {Kalogera}, \&
  {Bulik}}]{Belczynski02}
{Belczynski}, K., {Kalogera}, V., \& {Bulik}, T. 2002, \apj, 572, 407

\bibitem[{{Beniamini} \& {Piran}(2016)}]{Beniamini16}
{Beniamini}, P., \& {Piran}, T. 2016, \mnras, 456, 4089

\bibitem[{{Blaauw}(1961)}]{Blaauw61}
{Blaauw}, A. 1961, \bain, 15, 265

\bibitem[{{Blaes} {et~al.}(2002){Blaes}, {Lee}, \& {Socrates}}]{Blaes02}
{Blaes}, O., {Lee}, M.~H., \& {Socrates}, A. 2002, \apj, 578, 775

\bibitem[{{Bothwell} {et~al.}(2011){Bothwell}, {Kenicutt}, {Johnson}, {Wu},
  {Lee}, {Dale}, {Engelbracht}, {Calzetti}, \& {Skillman}}]{Bothwell11}
{Bothwell}, M.~S., {Kenicutt}, R.~C., {Johnson}, B.~D., {et~al.} 2011, \mnras,
  415, 1815

\bibitem[{{Brandt} \& {Podsiadlowski}(1995)}]{Brandt95a}
{Brandt}, N., \& {Podsiadlowski}, P. 1995, \mnras, 274, 461

\bibitem[{{Brandt} {et~al.}(1995){Brandt}, {Podsiadlowski}, \&
  {Sigurdsson}}]{Brandt95b}
{Brandt}, W.~N., {Podsiadlowski}, P., \& {Sigurdsson}, S. 1995, \mnras, 277,
  L35

\bibitem[{{Dominik} {et~al.}(2012){Dominik}, {Belczynski}, {Fryer}, {Holz},
  {Berti}, {Bulik}, {Mandel}, \& {O'Shaughnessy}}]{Dominik12}
{Dominik}, M., {Belczynski}, K., {Fryer}, C., {et~al.} 2012, \apj, 759, 52

\bibitem[{{Duch{\^e}ne} \& {Kraus}(2013)}]{Duchene13}
{Duch{\^e}ne}, G., \& {Kraus}, A. 2013, \araa, 51, 269

\bibitem[{{Gnedin} {et~al.}(2002){Gnedin}, {Zhao}, {Pringle}, {Fall}, {Livio},
  \& {Meylan}}]{Gnedin02}
{Gnedin}, O.~Y., {Zhao}, H., {Pringle}, J.~E., {et~al.} 2002, \apjl, 568, L23

\bibitem[{{Hale}(1994)}]{Hale94}
{Hale}, A. 1994, \aj, 107, 306

\bibitem[{{Ivanov} {et~al.}(2005){Ivanov}, {Polnarev}, \& {Saha}}]{Ivanov05}
{Ivanov}, P.~B., {Polnarev}, A.~G., \& {Saha}, P. 2005, \mnras, 358, 1361

\bibitem[{{Katz} \& {Dong}(2012)}]{Katz12}
{Katz}, B., \& {Dong}, S. 2012, ArXiv e-prints, arXiv:1211.4584

\bibitem[{{Kiseleva} {et~al.}(1996){Kiseleva}, {Aarseth}, {Eggleton}, \& {de La
  Fuente Marcos}}]{Kiseleva96}
{Kiseleva}, L.~G., {Aarseth}, S.~J., {Eggleton}, P.~P., \& {de La Fuente
  Marcos}, R. 1996, in Astronomical Society of the Pacific Conference Series,
  Vol.~90, The Origins, Evolution, and Destinies of Binary Stars in Clusters,
  ed. E.~F. {Milone} \& J.-C. {Mermilliod}, 433

\bibitem[{{Kobulnicky} {et~al.}(2014){Kobulnicky}, {Kiminki}, {Lundquist},
  {Burke}, {Chapman}, {Keller}, {Lester}, {Rolen}, {Topel}, {Bhattacharjee},
  {Smullen}, {Vargas {\'A}lvarez}, {Runnoe}, {Dale}, \&
  {Brotherton}}]{Kobulnicky14}
{Kobulnicky}, H.~A., {Kiminki}, D.~C., {Lundquist}, M.~J., {et~al.} 2014,
  \apjs, 213, 34

\bibitem[{{Kroupa}(2001)}]{Kroupa01}
{Kroupa}, P. 2001, \mnras, 322, 231

\bibitem[{{Kushnir} {et~al.}(2016){Kushnir}, {Zaldarriaga}, {Kollmeier}, \&
  {Waldman}}]{Kushnir16}
{Kushnir}, D., {Zaldarriaga}, M., {Kollmeier}, J.~A., \& {Waldman}, R. 2016,
  ArXiv e-prints, arXiv:1605.03839

\bibitem[{{Mandel}(2016)}]{Mandel16a}
{Mandel}, I. 2016, \mnras, 456, 578

\bibitem[{{Mandel} \& {de Mink}(2016)}]{Mandel16b}
{Mandel}, I., \& {de Mink}, S.~E. 2016, \mnras, 458, 2634

\bibitem[{{Martins} {et~al.}(2005){Martins}, {Schaerer}, \&
  {Hillier}}]{Martins05}
{Martins}, F., {Schaerer}, D., \& {Hillier}, D.~J. 2005, \aap, 436, 1049

\bibitem[{{Meynet} \& {Maeder}(2003)}]{Meynet03}
{Meynet}, G., \& {Maeder}, A. 2003, \aap, 404, 975

\bibitem[{Misner {et~al.}(1973)Misner, Thorne, \& Wheeler}]{MTW}
Misner, C., Thorne, K., \& Wheeler, J. 1973, Gravitation (W.H. Freeman and
  Company), 1--1279

\bibitem[{{Naoz}(2016)}]{Naoz16}
{Naoz}, S. 2016, ArXiv e-prints, arXiv:1601.07175

\bibitem[{{Naoz} {et~al.}(2016){Naoz}, {Fragos}, {Geller}, {Stephan}, \&
  {Rasio}}]{Naoz15}
{Naoz}, S., {Fragos}, T., {Geller}, A., {Stephan}, A.~P., \& {Rasio}, F.~A.
  2016, \apjl, 822, L24

\bibitem[{{Peters}(1964)}]{Peters64}
{Peters}, P.~C. 1964, Physical Review, 136, 1224

\bibitem[{{Rein} \& {Liu}(2012)}]{Rein12}
{Rein}, H., \& {Liu}, S.-F. 2012, \aap, 537, A128

\bibitem[{{Rein} \& {Spiegel}(2015)}]{Rein15}
{Rein}, H., \& {Spiegel}, D.~S. 2015, \mnras, 446, 1424

\bibitem[{{Repetto} \& {Nelemans}(2015)}]{Repetto15}
{Repetto}, S., \& {Nelemans}, G. 2015, \mnras, 453, 3341

\bibitem[{{Rodriguez} {et~al.}(2016{\natexlab{a}}){Rodriguez}, {Chatterjee}, \&
  {Rasio}}]{Rodriguez16a}
{Rodriguez}, C.~L., {Chatterjee}, S., \& {Rasio}, F.~A. 2016{\natexlab{a}},
  \prd, 93, 084029

\bibitem[{{Rodriguez} {et~al.}(2016{\natexlab{b}}){Rodriguez}, {Morscher},
  {Pattabiraman}, {Chatterjee}, {Haster}, \& {Rasio}}]{Rodriguez16b}
{Rodriguez}, C.~L., {Morscher}, M., {Pattabiraman}, B., {et~al.}
  2016{\natexlab{b}}, Physical Review Letters, 116, 029901

\bibitem[{{Sana} {et~al.}(2013){Sana}, {de Koter}, {de Mink}, {Dunstall},
  {Evans}, {H{\'e}nault-Brunet}, {Ma{\'{\i}}z Apell{\'a}niz},
  {Ram{\'{\i}}rez-Agudelo}, {Taylor}, {Walborn}, {Clark}, {Crowther},
  {Herrero}, {Gieles}, {Langer}, {Lennon}, \& {Vink}}]{Sana13}
{Sana}, H., {de Koter}, A., {de Mink}, S.~E., {et~al.} 2013, \aap, 550, A107

\bibitem[{{Sana} {et~al.}(2014){Sana}, {Le Bouquin}, {Lacour}, {Berger},
  {Duvert}, {Gauchet}, {Norris}, {Olofsson}, {Pickel}, {Zins}, {Absil}, {de
  Koter}, {Kratter}, {Schnurr}, \& {Zinnecker}}]{Sana14}
{Sana}, H., {Le Bouquin}, J.-B., {Lacour}, S., {et~al.} 2014, \apjs, 215, 15

\bibitem[{{Schaller} {et~al.}(1992){Schaller}, {Schaerer}, {Meynet}, \&
  {Maeder}}]{Schaller92}
{Schaller}, G., {Schaerer}, D., {Meynet}, G., \& {Maeder}, A. 1992, \aaps, 96,
  269

\bibitem[{{Seto}(2013)}]{Seto13}
{Seto}, N. 2013, Physical Review Letters, 111, 061106

\bibitem[{{Shapiro} \& {Teukolsky}(1983)}]{Shapiro83}
{Shapiro}, S.~L., \& {Teukolsky}, S.~A. 1983, {Black Holes, White Dwarfs and
  Neutron Stars} (New York: Wiley-Interscience)

\bibitem[{{Wegg}(2012)}]{Wegg12}
{Wegg}, C. 2012, \apj, 749, 183

\bibitem[{{Wen}(2003)}]{Wen03}
{Wen}, L. 2003, \apj, 598, 419

\bibitem[{{Willems} {et~al.}(2005){Willems}, {Henninger}, {Levin}, {Ivanova},
  {Kalogera}, {McGhee}, {Timmes}, \& {Fryer}}]{Williams05}
{Willems}, B., {Henninger}, M., {Levin}, T., {et~al.} 2005, \apj, 625, 324

\bibitem[{{Willems} {et~al.}(2004){Willems}, {Kalogera}, \&
  {Henninger}}]{Willems04}
{Willems}, B., {Kalogera}, V., \& {Henninger}, M. 2004, \apj, 616, 414

\end{thebibliography}
\appendix
\section{Algorithm for Selecting the Masses of Binary Stars Given an IMF and a Mass Ratio Distribution}
\label{binSelecAlg}

In this appendix we describe a method for picking a set of binary masses that satisfies a particular IMF and distribution of the mass ratio $\kappa$.  We assume that the IMF has an upper cutoff $m_{\rm max}$ and that the mass distribution of stars that are formed in binary systems with any mass ratio $\kappa$ is the same as the IMF.  This assumption cannot be satisfied for an arbitrary IMF, but it leads to no contradictions for the IMF assumed in Equation \eqref{NofM}.  We then apply this method to the specific case described in Section \ref{setup}, and derive Equations \eqref{alphaExact} and \eqref{leveleq}. 
\par
We pick the mass $m_1$ of the first star and the mass ratio $\kappa$ from the IMF and mass ratio distribution respectively.  Having picked $\kappa$, we only consider stars in systems with mass ratio $\kappa$.  We define $\alpha(m, \kappa)$ such that a fraction $\alpha(m, \kappa)$ of stars with mass in the range $[m, m + dm]$ are paired with stars with mass in the range $[m/\kappa, (m+dm)/\kappa]$.  We then note that $1-\alpha(m/\kappa, \kappa)$ of stars in the range $[m/\kappa, (m + dm)/\kappa]$ are paired with stars in the range $[m, m + dm]$.  Let there be $N_\kappa(m)dm$ stars with mass in the range $[m, m + dm]$ in binaries with mass ratio $\kappa$.  These observations allow us to write
\begin{equation}
\label{alphaEquation}
\alpha(m, \kappa)N_\kappa(m)dm = [1-\alpha(m/\kappa, \kappa)] N_\kappa(m/\kappa)dm/\kappa.
\end{equation}
\par
If $m/\kappa > m_{\rm max}$, we can set $\alpha(m, \kappa) = 0$.   Now, let 
\begin{equation}
 \label{leveleqappendix}
L = \left \lfloor \frac{\ln{m} - \ln{m_{\rm max}}}{\ln{\kappa}}\right \rfloor.
\end{equation}
This has the property that $\kappa^{L+1} m_{\rm max} < m \leq \kappa^L m_{\rm max}$.  We see therefore, from repeated application of Equation \eqref{alphaEquation}, that $\alpha(m, \kappa)$ is the $L^{\rm th}$ term in the sequence where $x_0 = 0$ and 
\begin{equation}
\label{recurrenceRelationGeneral}
x_{i+1}N_\kappa(m/\kappa^{L-i-1}, \kappa) = (1-x_i)N_\kappa(m/\kappa^{L-i}, \kappa)/\kappa.
\end{equation}
We then let the mass of the second star be $\kappa m_1$ with probability $1-\alpha(m_1, \kappa)$ and $m_1/\kappa$ with probability $\alpha(m-1, \kappa)$.
\par
For the specific case of the form $N_\kappa(m) \sim m^{-2.3}$ discussed in Section \ref{setup}, Equation \eqref{recurrenceRelationGeneral} simplifies to 
\begin{equation}
\label{recurrenceRelation}
x_{i+1} = [1-x_i] \kappa^{1.3}.
\end{equation}
It is trivial to show via induction that the $L^{\rm th}$ term in the sequence defined by Equation \eqref{recurrenceRelation} and the initial condition $x_0 = 0$ is given by Equation \eqref{alphaExact}.  Therefore $\alpha(m, \kappa)$ is given by Equation \eqref{alphaExact}, with $L$ given by Equation \eqref{leveleq}.

\section{Verification of our Gravitational-Wave Algorithm}
\label{verification}
 \begin{figure}
\centering
\includegraphics[width=.8\textwidth]{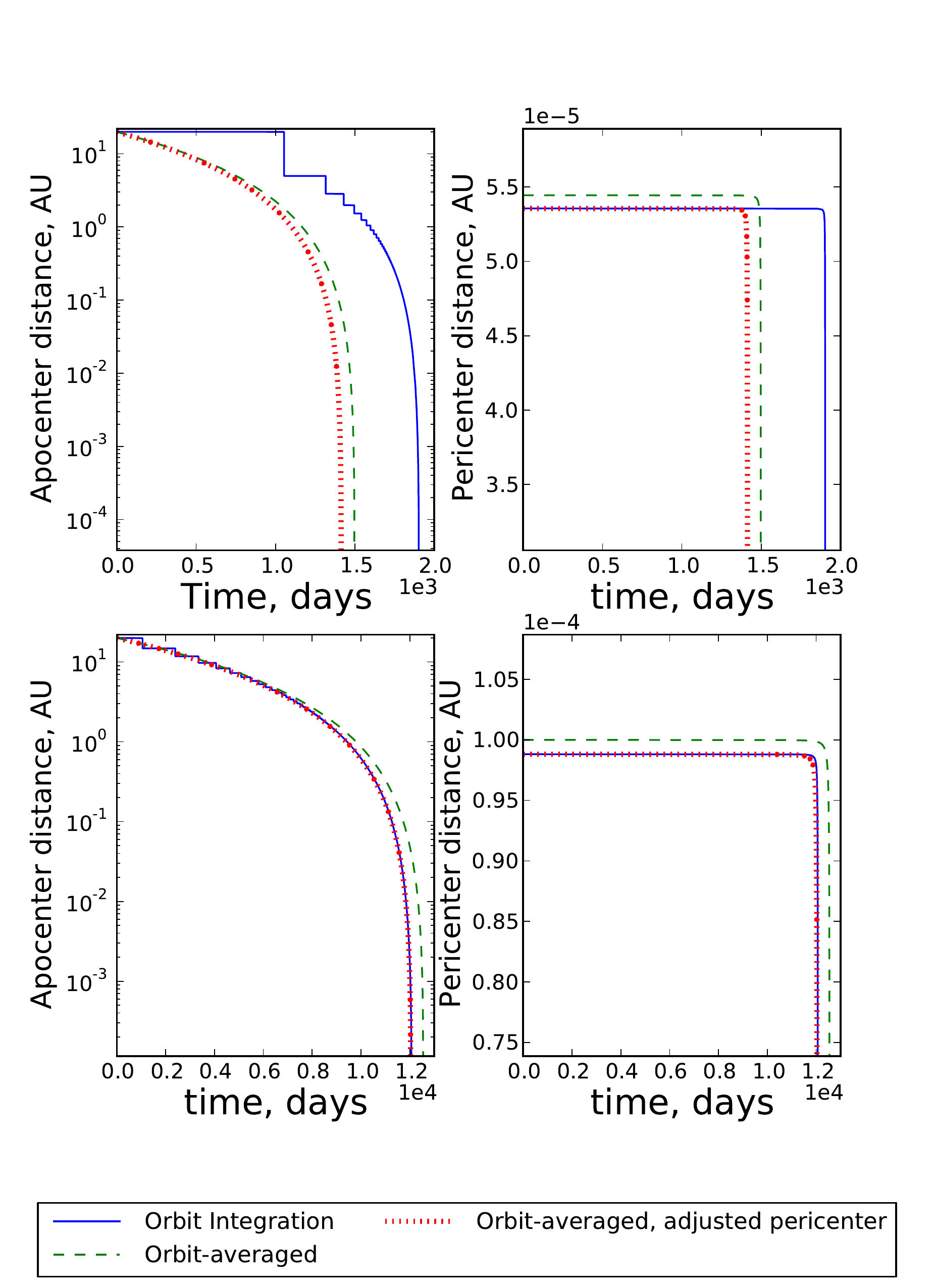}
\caption{Time evolution of apocenter distance and pericenter distance in our simulations (blue solid line) compared with the orbit-averaged quadrupole approximation.  The green dashed curve corresponds to the orbit-averaged case with the same initial $a$ and $e$, and the red dotted curve to the orbit-averaged case of a system with the same initial pericenter and apocenter distance.    }
\label{Inspiral1}
\vspace{-.05cm}
\end{figure} 

 \begin{figure}
\centering
\includegraphics[width=.8\textwidth]{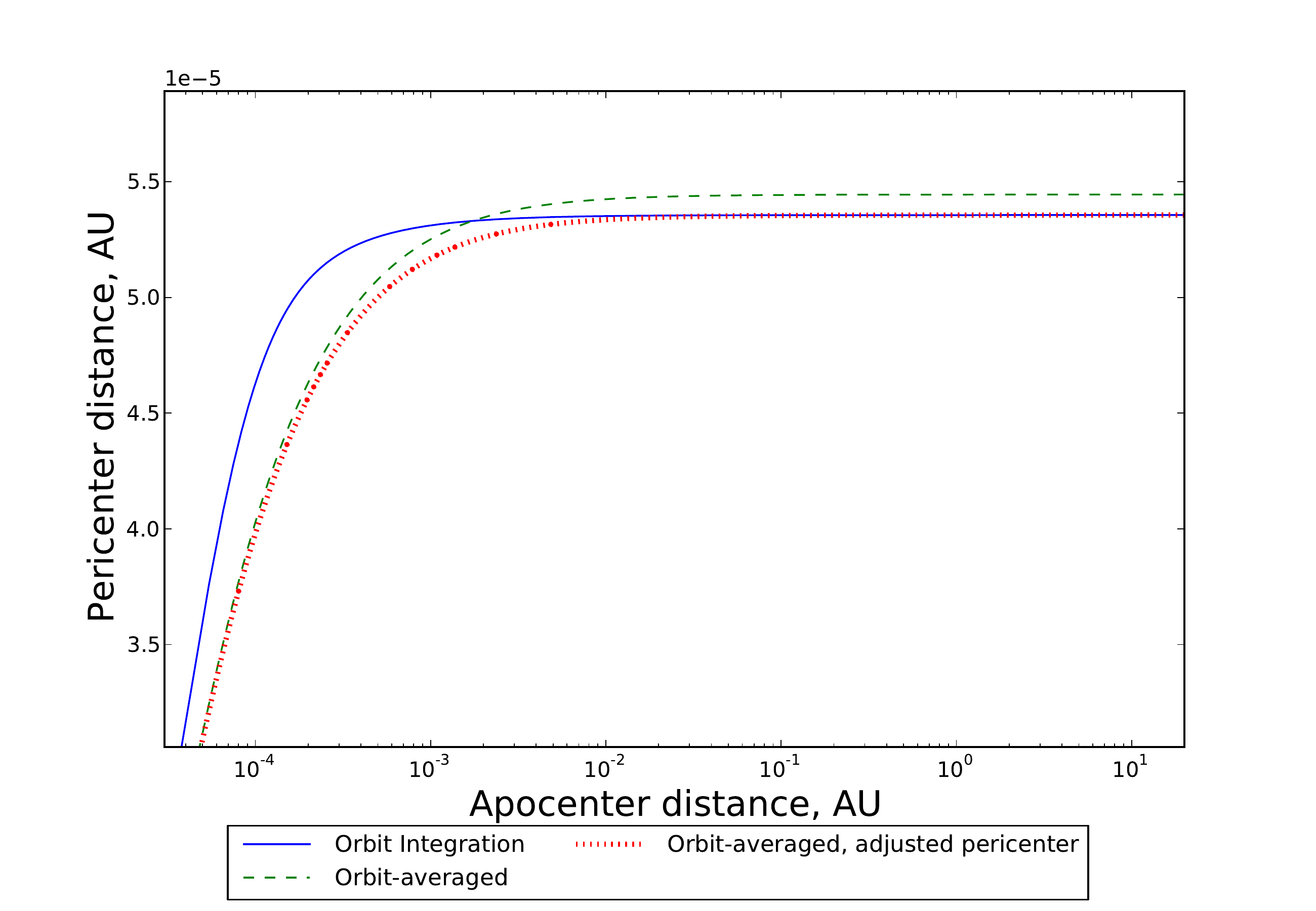}
\caption{Pericenter distance vs. apocenter distance in our simulations (blue solid curve) compared with the time averaged quadrupole approximation.  The green dashed curve corresponds to the orbit-averaged case with the same initial value of $a$ and $e$, and the red dotted curve to the orbit-averaged case of a system with the same initial pericenter and apocenter distance.  }
\label{Inspiral2}
\vspace{-.05cm}
\end{figure} 

In this appendix we compare our orbital inspiral simulations to the orbit-averaged quadrupole approximation.  We compare the evolution of the apocenter and pericenter distance because the traditional orbital elements are poorly-behaved when the orbits are highly eccentric and have small pericenter distances, due to the extra potential $U_{\rm GR}$ that we added to reproduce the apsidal precession due to general relativity.  In order for the orbital elements to not be changed significantly by the extra term in the potential, we require $U_{\rm GR}/E_b \ll 1$, which implies 
\begin{equation}
\frac{3R_{\rm EH}a}{R_{ij}^2} \ll 1,
\end{equation}
where $R_{\rm EH} = 2GM/c^2$ is the Schwarzschild radius (radius of the event horizon) corresponding to the sum of the masses, $R_{ij}$ is the instantaneous separation, and $E_b$ is the binding energy of the binary orbit.  Note that for highly eccentric orbits, just the restriction $R_{\rm EH} \ll R_{ij}$ is insufficient.
\par
To circumvent this problem with the traditional orbital elements, we numerically calculate the pericenter and apocenter distance of an orbit in the potential $U = U_{\rm GR} + U_{\rm Kep}$ (where $U_{\rm Kep}$ is the Keplerian potential) given the positions, velocities and masses of the two components.  
\par
Figure \ref{Inspiral1} shows the evolution of the apocenter and pericenter distance of two simulated orbits.  In the upper panels, the system was started at apocenter with $a = 10$ AU, $e$ = 0.9999946.  This yields a value of $a(1-e)$ equal to $q_{\rm crit}$ calculated in Equation \eqref{qcrit}, assuming that $a_{\rm out} = 50$ AU, and $m_3 = 20 M_\odot$.  As discussed above, the true pericenter distance will be smaller by a few percent (the difference between the green and red curves in the top right panel).  In the lower panels, the system was started at apocenter with $a_{\rm in}$ = 10 AU, $e_{\rm in}$ = 0.99999.
\par
In both cases, the blue curves are the result from our orbit integrations.  The pericenter distance is not equal to $a_{\rm in} (1-e_{\rm in})$ because we have modified the potential.  The green curve is the result from the orbit-averaged quadrupole formula starting with the same semi-major axis and eccentricity as the simulated system.  The red curve is the result from the orbit-averaged quadrupole formula starting with the same apocenter and pericenter distance as the simulated system.  
\par
In the top panels, both the red and green curves are very different from the blue because initially there is a large drop in semi-major axis in one pericenter passage.  This makes the orbit-averaged approximation a poor one; the true system takes longer to merge because it has to wait $1/2$ of an orbit to radiate any energy.  We see that the red and green curves only differ by a few percent, implying that the difference in pericenter is not very significant.  In the bottom panels, the discrete jumps in $a$ are smaller, so the red curve overlaps the blue one almost perfectly.  
\par
Figure \ref{Inspiral2} shows the evolution of $q_{\rm in}$ vs. $a_{\rm in}$.  This is made for the system simulated in the upper two panels of Figure \ref{Inspiral1}  We see that deviations between our method (blue) and the orbit-averaged quadrupole approximation with the same pericenter distance (red) are only significant for $a_{\rm in} < 0.05$ AU, long after the orbit is decoupled from the outer companion.
\par
We conclude from these tests that while our method is not perfect, it is substantially closer to reality than an orbit-averaged approximation.  Given the substantial uncertainties about many of the astrophysical parameters in this paper, we believe that deviations between our orbit integrations and the true orbits are not a substantial source of error.

\end{document}